\documentclass[%
 amsmath,amssymb,
 aps,pre
]{revtex4-1}

\usepackage{graphicx}
\usepackage{dcolumn}
\usepackage{bm}
\usepackage{amssymb, amsmath, amsthm, amsfonts}
\usepackage{subcaption} 
\usepackage{gensymb}
\usepackage{color}
\usepackage{mathtools} 

\usepackage{graphicx} 

\usepackage{tikz} 

\usepackage{hyperref}
\usepackage[mathlines]{lineno}

\DeclareMathSizes{12}{12}{7}{0.5} 

\newcommand{\Arrow}[1]{%
\parbox{#1}{\tikz{\draw[->](0,0)--(#1,0);}}
}

\newcommand*{\transp}[2][-3mu]{\ensuremath{\mskip1mu\prescript{\smash{\mathbf T \mkern#1}}{}{\mathstrut#2}}}%

\begin{document}



\title{Explicit and implicit network connectivity: Analytical formulation and application to transport processes}

\author{Enrico Ser-Giacomi}
\email{enrico.sergiacomi@gmail.com}
\affiliation{Department of Earth, Atmospheric and Planetary Sciences, Massachusetts Institute of Technology, 54-1514 MIT, Cambridge, MA 02139, USA.}

\author{T\'{e}rence Legrand}
\affiliation{Aix Marseille Univ., University of Toulon, CNRS, IRD, Mediterranean Institute of Oceanography (UMR 7294), Marseille, France.}

\author{Ismael Hernandez-Carrasco}
\affiliation{Mediterranean Institute for Advances Studies (IMEDEA, UIB-CSIC), Mallorca (Spain)}

\author{Vincent Rossi}
\affiliation{Aix Marseille Univ., Universite de Toulon, CNRS, IRD, Mediterranean Institute of Oceanography (UMR 7294), Marseille, France.$\,$}

\begin{abstract}

Connectivity is a fundamental structural feature of a network that determines the outcome of any dynamics that happens on top of it. However, an analytical approach to obtain connection probabilities between nodes associated to paths of different lengths is still missing. Here, we derive exact expressions for random-walk connectivity probabilities across any range of numbers of steps in a generic temporal, directed and weighted network. This allows characterizing explicit connectivity realized by causal paths as well as implicit connectivity related to motifs of three nodes and two links called here pitchforks. We directly link such probabilities to the processes of tagging and sampling any quantity exchanged across the network, hence providing a natural framework to assess transport dynamics. Finally, we apply our theoretical framework to study ocean transport features in the Mediterranean Sea. We find that relevant transport structures, such as fluid barriers and corridors, can generate contrasting and counter-intuitive connectivity patterns bringing novel insights into how ocean currents drive seascape connectivity.

\end{abstract}

\maketitle




\section{Introduction}

Connectivity is a key feature of network's structure \cite{boccaletti2006complex,holme2012temporal} that determines how strongly and quickly different nodes can be linked by consecutive edges \cite{watts1998collective, estrada2008communicability, lentz2013unfolding}. Indeed, for any dynamics running over a network, connectivity strongly influences the temporal and spatial evolution of the associated processes and patterns \cite{barrat2008dynamical}. This has been proven in several contexts such as epidemic or information spreading \cite{moreno2002epidemic,ulanowlcz1990symmetrical}, biological interactions \cite{garlaschelli2003universal}, neural networks \cite{rubinov2010complex}, social systems \cite{kempe2005influential} and fluid transport \cite{ser2015flow}. Globally, connectivity is determined by topological properties of the network: link density, degree and weight distributions, clustering, modularity, reciprocity, etc. However, these metrics describe statistical features of the network and do not inform about local patterns of connectivity between specific pairs of nodes \cite{watts1998collective,newman2012communities}. 

The conventional approach to characterize pair-wise connectivity consists in studying random walks and their trajectories. In fact, random walkers can be seen as agents that navigate through the network drawing paths between pairs of nodes \cite{lovasz1993random, newman2005measure, masuda2017random}. Each of these pathways can be thus defined by the sequence of nodes visited by a random walker and, by multiplying the node-to-node single-step transition probabilities, one can obtain the probability of occurrence of any of them \cite{brockmann2013hidden, ser2015most, gautreau2007arrival, huntsman2018markov}.

In this way, connectivity can be characterized within a solid probabilistic framework. Moreover, when a given quantity, such as people \cite{barrat2004architecture}, fluid \cite{ser2015flow}, goods \cite{serrano2003topology} or information \cite{pastor2007evolution}, is transported across the network, random walker transition probabilities can be related to fractions of exchanged quantities between node pairs. More concretely, this means that it is possible to calculate the probability that an amount of quantity that has been tagged or sampled in a given node will reach another specific destination node forward- or backward-in-time, respectively. As a result, random walks can also mimic transport, dispersion and mixing processes across a network \cite{ser2015most, ser2015dominant}. Eventually, this could permit to rigorously establish a quantitative link between the structural features of a network and the dynamics of any transported quantity across it. Such connection would also be relevant in temporal networks, especially under mixing regimes in which network connectivity patterns and random walks unfold on comparable time scales \cite{perra2012random,starnini2012random}. 

However, to our knowledge, analytical expressions for connectivity probabilities between any pair of nodes that take into account connections realized by paths of different lengths (i.e. paths composed of different number of steps) in a ``cumulated'' manner are, to our knowledge, still lacking. Indeed, while  the probabilities of connection realized by paths of the same lengths (i.e. imposing a prescribed number of steps) are readily obtained with simple matrix products, the cumulated probabilities across generic ranges of path lengths (i.e. across different numbers of steps) has not been derived yet. This is mainly due to the fact that connection events between two nodes realized by paths of different lengths are not mutually-exclusive from a probabilistic point of view, making the calculations to obtain them quite convoluted. It is worth noting that this shortcoming holds for both static and temporal network. As such, the current approach to study pairwise connectivity is through Monte Carlo numerical simulations. Specifically, it consists in releasing large numbers of random walkers in a given starting node and in estimating the connection probability with any other destination node from the proportion of walkers that ended up there after a given number of steps.

\begin{figure}
    \includegraphics[width=10cm]{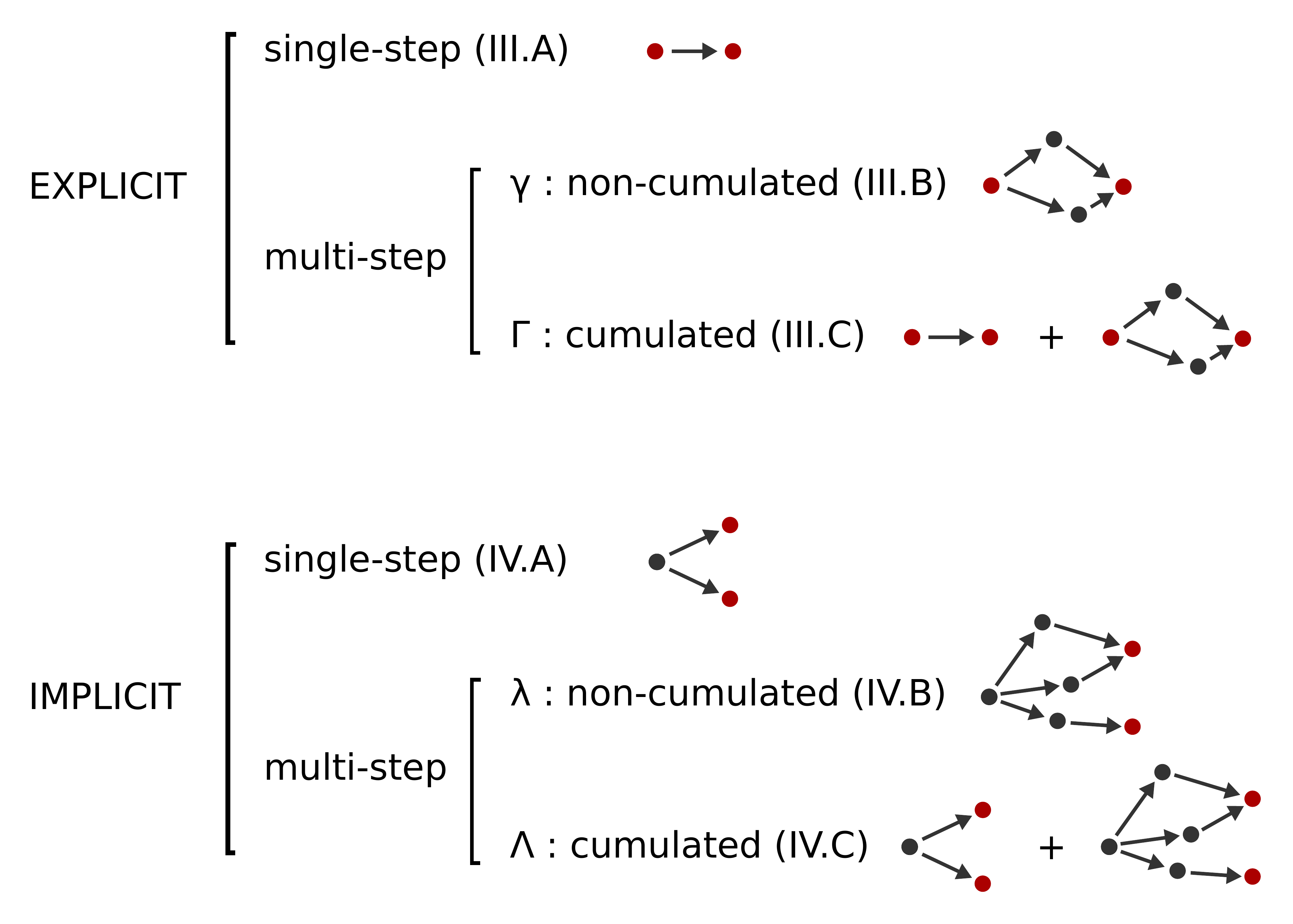} 
    \caption{Sketch of the different connectivity probabilities considered. The sections introducing each quantity are indicated within brackets while greek letters correspond to their mathematical expressions.} \label{fig:schemaintro}
\end{figure}

Moreover, also the concept of connectivity by itself could be extended. Indeed, the \emph{explicit} connectivity probabilities described above are conceptually associated with the pathway of a random walker that joins two nodes, symbolizing a kind of ``parent-child'' relationship between starting and ending node. Nevertheless, we can also be interested in looking contemporaneously at the entire network in a synoptic fashion. This is the case, for instance, when modeling a transport or spreading phenomena on a network \cite{wang2017unification} or when tracking differentiation across a phylogenetic tree \cite{woese2000interpreting}. In such processes, each pair of nodes could be simultaneously influenced by a third node (or more than one) and such ``sibling-sibling'' relationships can determine similarities between nodes pairs that we could regard as a form of \emph{implicit} connectivity. These connectivity patterns, at one step, are realized by a particular kind of three-nodes motif, here called \emph{pitchfork}, composed of a node acting as common source (or destination) for two other nodes (see Fig. \ref{fig:schemaintro}). Examples of such kind of interactions can be found in ecological networks  when two species compete for the same resource \cite{gracia2018joint} or in social systems when two agents  are both influenced by a third one \cite{chen2012identifying}. Thus, implicit connections associated with pitchforks are in this sense complementary to the aforementioned standard explicit patterns and, despite being mostly overlooked, could play a major role in determining network dynamics. 

In this paper, we derive exact analytical expressions for explicit and implicit random-walk connectivity probabilities across any range of numbers of steps in a generic temporal, directed and weighted network. First, in Section \ref{sec:random} we set the theoretical background and delineate the relationships between random walk transition probabilities and the transport dynamics of a given quantity across the network. In Sections \ref{sec:direct} and \ref{sec:implicit}, we introduce the concept of \emph{cumulated} connectivity that permits to calculate connection probabilities not only for a fixed number of steps but also across an arbitrary range of possible numbers of steps, allowing probability values to eventually saturate toward an asymptotic value. Such approach is adopted to provide exact formulas for: (i) explicit connectivity patterns associated with causal paths among two nodes and (ii) implicit connectivity patterns realized by multistep pitchforks (see a summary of the different connectivity patterns in Fig. \ref{fig:schemaintro}). Moreover, if a given quantity $Q$ is transported across the network, we can relate random walk probabilities to processes of tagging and sampling such quantity in specific nodes of the network. This allows linking the probabilistic view of connectivity with an interpretation in terms of transport and diffusion. In Section \ref{sec:toy}, we calculate connection probabilities for two simple networks and we numerically confirm our analytical results highlighting significant differences between static and temporal network connectivity. In Section \ref{sec:medsea}, we further apply our theoretical approach to characterize connectivity probabilistic features of a network describing the transport of surface water masses across the Mediterranean Sea \cite{ser2015flow, miron2017lagrangian, donner2019characterizing, banisch2019network}. From probabilistic estimations of connectivity we provide both specific site-to-site and global basin-scale statistics. We find very relevant differences among explicit and implicit probabilities and across different ranges of number of steps. We also show that such probabilities, in average, saturate to different, non-trivial values. Finally, we discuss the implications of such results.

\section{Random walks and transport processes}
\label{sec:random}

\subsection{Network adjacency matrix and its normalizations}

We consider a generic directed, weighted and temporal network of $N$ nodes. Hence, each of its links is directed and characterized by a positive weight that measures the 'intensity' of the connection realized between two nodes. Moreover, due to the temporal character of the network, such weights can change in time. Given a discrete time sequence $\bigl\{t_0, t_{1}, \, ...\, , t_{M-1}, t_M \bigl\}$, the time-dependent structure of the network can be thus described by a set of adjacency matrices in which each element $\mathbf{A}^{t_l \,\Arrow{.15cm}\, t_{l+1}}_{ij}$ is the weight of the link from node $i$ to node $j$ during the time interval $[t_{l},t_{l+1}]$. For convention, links are hereafter established forward-in-time across different layers representing consecutive discrete times \cite{kim2012temporal}.

We define the out-strength and in-strength of node $i$ as:
\begin{align}
S^{O}_{i}(t_l \,\Arrow{.15cm}\, t_{l+1}) &= \sum_j \mathbf{A}^{t_l \,\Arrow{.15cm}\, t_{l+1}}_{ij},   \label{eq:outstreng} \\
S^{I}_{i}(t_l \,\Arrow{.15cm}\, t_{l+1}) &= \sum_j \mathbf{A}^{t_l \,\Arrow{.15cm}\, t_{l+1}}_{ji}. \label{eq:instreng}
\end{align}
Assuming that $S^{O}$ and $S^{I}$ are always positive, two normalizations for the matrix $\mathbf{A}^{t_l \,\Arrow{.15cm}\, t_{l+1}}_{ij}$ are possible:
\begin{align}
\mathbf{F}^{t_l \,\Arrow{.15cm}\, t_{l+1}}_{ij} &= \frac{\mathbf{A}^{t_l \,\Arrow{.15cm}\, t_{l+1}}_{ij}}{S^{O}_{i}(t_l \,\Arrow{.15cm}\, t_{l+1})},   \label{eq:outnormmat} \\
\mathbf{B}^{t_{l+1} \,\Arrow{.15cm}\, t_l}_{ji} &= \frac{\mathbf{A}^{t_l \,\Arrow{.15cm}\, t_{l+1}}_{ij}}{S^{I}_{j}(t_l \,\Arrow{.15cm}\, t_{l+1})}, \label{eq:innormmat}
\end{align}
obtaining the following conservation conditions: $\sum_{j} \mathbf{F}^{t_l \,\Arrow{.15cm}\, t_{l+1}}_{ij} = 1$ and $\sum_{i} \mathbf{B}^{t_{l+1} \,\Arrow{.15cm}\, t_l}_{ji} = 1$.

\subsection{Random walk transition probabilities and transport dynamics}
\label{subsec:transwalk}

Once the adjacency matrices of the network  $\mathbf{A}^{t_l \,\Arrow{.15cm}\, t_{l+1}}$ are normalized, a random walk can be defined on it. Indeed, in the $[t_l;t_{l+1}]$ time interval, $\mathbf{F}^{t_l \,\Arrow{.15cm}\, t_{l+1}}_{ij}$ is the forward-in-time transition probability for a random walker to jump from node $i$ to $j$ while $\mathbf{B}^{t_{l+1} \,\Arrow{.15cm}\, t_l}_{ji}$ is the backward-in-time transition probability to go from $j$ to $i$. Hence, the direction of the links is always associated with the forward-in-time direction but still, for a given link, we are able to define both the forward- and backward-in-time transition probabilities. If we assume a Markovian dynamics, the probability for a random walker to visit a given sequence of nodes will be given by the product of the associated single-step transition probabilities.

If link weights can be associated with a generic transported quantity $Q$ across the network, random walk transition probabilities can be related to processes of tagging and sampling the transported quantity. Indeed, imagining to tag a portion of $Q$ inside $i$ at $t_l$, $\mathbf{F}^{t_l \,\Arrow{.15cm}\, t_{l+1}}_{ij}$ is the probability that such tagged quantity will arrive to $j$ at $t_{l+1}$. Consequently, $\mathbf{B}^{t_{l+1} \,\Arrow{.15cm}\, t_l}_{ji}$ is the probability of sampling a portion of $Q$ in $j$ at $t_{l+1}$ that was in $i$ at $t_l$. Pushing forward this analogy, we can quantify the fraction of transported quantity between the pair of nodes $i,j$ in the time interval $[t_l;t_{l+1}]$ by means of transition probabilities \cite{ser2015most,ser2015dominant}. Indeed, $\mathbf{F}^{t_l \,\Arrow{.15cm}\, t_{l+1}}_{ij}$ is the fraction of $Q$ present in $i$ at $t_l$ that arrives to $j$ at $t_{l+1}$. Similarly, $\mathbf{B}^{t_{l+1} \,\Arrow{.15cm}\, t_l}_{ji}$ is the fraction of $Q$ present in $j$ at $t_{l+1}$ that was in $i$ at $t_l$.

\subsection{Paths in temporal weighted networks}

We denote a path $\mu$ of $M$-steps between nodes $i$ and $j$ as a $(M+1)$-tuple $\bigl\{i, k_{1},\, ...\, , k_{M-1}, j \bigl\}$ corresponding to the sequence of nodes visited by a random walker at times $\bigl\{t_0, t_{1}, \, ...\, , t_{M-1}, t_M \bigl\}$. 

Thus, assuming a Markov process, the forward-in-time probability for a random walker to take the $M$-steps path $\mu$ under the condition of starting in $i$ and ending in $j$ is \cite{ser2015most,ser2015dominant,huntsman2018markov,koltai2018large}:
\begin{equation}\label{eq:mostprobpathprobfor}
\mathbf{F}^{t_0 \,\Arrow{.15cm}\, t_1}_{ik_{1}} \, \mathbf{F}^{t_1 \,\Arrow{.15cm}\, t_2}_{k_{1}k_{2}}  \,...\,
\mathbf{F}^{t_{M-2} \,\Arrow{.15cm}\, t_{M-1}}_{k_{M-2}k_{M-1}} \, \mathbf{F}^{t_{M-1} \,\Arrow{.15cm}\, t_M}_{k_{M-1}j}.
\end{equation}
Conversely, the backward-in-time probability to take the $M$-steps path $\mu$ under the condition of starting in $j$ and ending in $i$ is:
\begin{equation}\label{eq:mostprobpathprobback}
\mathbf{B}^{t_{M} \,\Arrow{.15cm}\, t_{M-1}}_{jk_{M-1}} \,\mathbf{B}^{t_{M-1} \,\Arrow{.15cm}\, t_{M-2}}_{k_{M-1}k_{M-2}}  \,...\,
\mathbf{B}^{t_{2} \,\Arrow{.15cm}\, t_{1}}_{k_{2}k_1} \,\mathbf{B}^{t_{1} \,\Arrow{.15cm}\, t_{0}}_{k_{1}i} .
\end{equation}

Note that, due to the temporal dependence of the network, the above probabilities depend not only on the number of steps (as in the static case) but also on the specific initial or final time considered.

\section{Explicit connectivity}
\label{sec:direct}

In this Section we provide exact analytical expressions for random walk probabilities associated with paths. Depending on the number of steps considered, we can define single-step ($M=1$) or multistep ($M>1$) connectivity. First, we introduce connectivity for the case of a fixed number of steps $M$ (non-cumulated connectivity). Then, we extend this conventional concept by considering connections occurring over a given range of number of steps spanning 1 to $M$ (cumulated connectivity). Hence, the non-cumulated connectivity is associated with the probability that a random walker joins two nodes in a specific number of steps. Note that, for the temporal case, since the network is time-dependent, we should also specify the initial time. This probability does not include the possibility of reaching the destination node before or after the exact number of steps chosen. Cumulated connectivity overcomes this limitation by considering the probability that a random walker reaches the destination in an arbitrary number of steps as long as it is comprised within a given range of numbers of steps. When a generic quantity $Q$ is transported across the network, we can also find a relation between the above probabilities and portions of $Q$ (see Section \ref{subsec:transwalk}).

\subsection{Single-step explicit connectivity}

Single-step explicit connectivity is associated directly with the elements of the $\mathbf{F}$ and $\mathbf{B}$ matrices (see Section \ref{subsec:transwalk}). Considering the $[t_{l},t_{l+1}]$ time interval, we define the \emph{single-step explicit connectivity} calculated forward-in-time from node $i$ to $j$ as:
\begin{equation}
\boldsymbol{\gamma}^{\, f} (t_l,t_{l+1}) = \mathbf{F}^{t_l \,\Arrow{.15cm}\, t_{l+1}}_{ij} .
\end{equation}
Similarly, we define the backward-in-time \emph{single-step explicit connectivity} as: 
\begin{equation}
\boldsymbol{\gamma}^{\, b} (t_l,t_{l+1}) = \mathbf{B}^{t_{l+1} \,\Arrow{.15cm}\, t_{l}}_{ij} .
\end{equation}
If some generic quantity $Q$ is transported across the network and one tags an amount of it that is present in node $i$ at time $t_l$, the probability that will arrive in node $j$ at time $t_{l+1}$ is exactly $\mathbf{F}^{t_l \,\Arrow{.15cm}\, t_{l+1}}_{ij}$. Analogously, if one samples an amount of $Q$ in $i$ at time $t_{l+1}$, the probability that was in $j$ at time $t_l$ is $\mathbf{B}^{t_{l+1} \,\Arrow{.15cm}\, t_{l}}_{ij}$.

\subsection{Non-cumulated multistep explicit connectivity}

To obtain the total probability of connection among any given pair of nodes in exactly $M$ steps, we need to sum the probability of each of the paths that connect that pair. Hence, using the Chapman-Kolmogorov equation, we define, given a fixed number of steps $M$, the \emph{non-cumulated multistep explicit connectivity} calculated forward-in-time as: 
\begin{equation}\label{eq:multist_for}
\boldsymbol{\gamma}^{\, f} (t_0,t_M) = \, \mathbf{F}^{t_0 \,\Arrow{.15cm}\, t_1} \, \mathbf{F}^{t_1 \,\Arrow{.15cm}\, t_2} \,...\, 
\mathbf{F}^{t_{M-2} \,\Arrow{.15cm}\, t_{M-1}} \, \mathbf{F}^{t_{M-1} \,\Arrow{.15cm}\, t_M}   \, .
\end{equation} 
Similarly, we define the \emph{non-cumulated multistep explicit connectivity} calculated backward-in-time :
\begin{equation}\label{}
\boldsymbol{\gamma}^{\, b} (t_0,t_M) = \, \mathbf{B}^{t_{M} \,\Arrow{.15cm}\, t_{M-1}} \,\mathbf{B}^{t_{M-1} \,\Arrow{.15cm}\, t_{M-2}}  \,...\,
\mathbf{B}^{t_{2} \,\Arrow{.15cm}\, t_{1}} \,\mathbf{B}^{t_{1} \,\Arrow{.15cm}\, t_{0}}  \,.
\end{equation}
In both definitions above we used the fact that summing probabilities over all the paths corresponds to performing the matrix product of the associated adjacency matrices. Therefore, $\boldsymbol{\gamma}^{\, f}$ is a matrix whose element $i-j$ is the probability for a random walker to reach $j$ starting from $i$ after $M$-steps forward-in-time. Similarly, $\boldsymbol{\gamma}^{\, b}$ is a matrix whose element $i-j$ is the probability for a random walker to reach $j$ starting from $i$ after $M$-steps backward-in-time. It is straightforward to prove that the matrix elements of both $\boldsymbol{\gamma}^{\, f}$ and $\boldsymbol{\gamma}^{\, b}$ are always bounded in between 0 and 1.

\subsection{Cumulated multistep explicit connectivity}\label{sec:cummultiexplicit}
We now consider the case of multiple numbers of steps together to introduce the novel concept of cumulated connectivity (see Fig. \ref{fig:multistepfig}). We still refer to the discrete time sequence  $\bigl\{t_0, t_{1}, \, ...\, , t_{M-1}, t_M \bigl\}$ introduced before and we provide the probability for a random walker to connect two nodes in a finite range of possible number of steps. In the forward-in-time case, the initial time $t_0$ is fixed and the number of steps increases progressively ending up at larger $t_l$'s. Backward-in-time, we instead end always at $t_M$ but starting from decreasing $t_l$'s while increasing the number of steps. Without loss of generality, we consider in the following multistep connectivity realized in a range of number of steps comprised between 1 and $M$.

\subsubsection{Deriving up to 3-steps cumulated multistep explicit connectivity}
Let's start considering the forward-in-time connectivity. Keeping fixed the initial time $t_0$, we focus on a starting node $i$ and a destination node $j$ and we aim to find an expression for the union of the multistep explicit probabilities of 1, 2 and 3 steps. We start defining the following three events:
\begin{itemize}
\item $A$: reaching $j$ from $i$ in 1 step,
\item $B$: reaching $j$ from $i$ in 2 steps,
\item $C$: reaching $j$ from $i$ in 3 steps.
\end{itemize}
Since the events connecting $i$ to $j$ in different numbers of steps are not mutually-exclusive, we cannot obtain the three-events union probability BY simply summing their individual probabilities. Such union probability, which will be used to define the cumulated multistep explicit connectivity, can be written as:
\begin{widetext}
\begin{align}
 &P(A \cup B \cup C)_{ij} = P(A) + P(B) + P(C) + P(A \cap B \cap C) - P(A \cap B)- P(A \cap C)- P(B \cap C) = \nonumber \\
 &= P(A) + P(B) + P(C) + P(A)P(B|A)P(C|A \cap B) - P(A)P(B|A) - P(A)P(C|A) - P(B)P(C|B) \,. \label{eq:prob_union}
\end{align}
\end{widetext}
Following Eq. (\ref{eq:multist_for}), the multistep connectivity probabilities of the events $A$, $B$ and $C$ are:
\begin{align}
& P(A)=\mathbf{F}^{t_0 \,\Arrow{.15cm}\, t_1}_{ij} \,, \label{eq:proba}\\
& P(B)=\sum_{k} \mathbf{F}^{t_0 \,\Arrow{.15cm}\, t_1}_{ik} \, \mathbf{F}^{t_1 \,\Arrow{.15cm}\, t_2}_{kj} \,,\label{eq:probb}\\
& P(C)=\sum_{k,l} \mathbf{F}^{t_0 \,\Arrow{.15cm}\, t_1}_{ik} \, \mathbf{F}^{t_1 \,\Arrow{.15cm}\, t_2}_{kl} \, \mathbf{F}^{t_2 \,\Arrow{.15cm}\, t_3}_{lj} \,. \label{eq:probc}
\end{align}
Using Eqs. (\ref{eq:proba},\ref{eq:probb},\ref{eq:probc}) into Eq.(\ref{eq:prob_union}) we obtain:
\begin{align}
 &P(A \cup B \cup C)_{ij} = \mathbf{F}^{t_0 \,\Arrow{.15cm}\, t_1}_{ij} \,
 +\, \sum_{k} \mathbf{F}^{t_0 \,\Arrow{.15cm}\, t_1}_{ik} \, \mathbf{F}^{t_1 \,\Arrow{.15cm}\, t_2}_{kj} \,
 +\, \sum_{k,l} \mathbf{F}^{t_0 \,\Arrow{.15cm}\, t_1}_{ik} \, \mathbf{F}^{t_1 \,\Arrow{.15cm}\, t_2}_{kl} \, \mathbf{F}^{t_2 \,\Arrow{.15cm}\, t_3}_{lj} \,
 +\, \mathbf{F}^{t_0 \,\Arrow{.15cm}\, t_1}_{ij} \, \mathbf{F}^{t_1 \,\Arrow{.15cm}\, t_2}_{jj} \, \mathbf{F}^{t_2 \,\Arrow{.15cm}\, t_3}_{jj} \,- \nonumber \\
 &-\, \mathbf{F}^{t_0 \,\Arrow{.15cm}\, t_1}_{ij} \, \mathbf{F}^{t_1 \,\Arrow{.15cm}\, t_2}_{jj} \,
 -\,  \sum_{l}\mathbf{F}^{t_0 \,\Arrow{.15cm}\, t_1}_{ij} \, \mathbf{F}^{t_1 \,\Arrow{.15cm}\, t_2}_{jl} \, \mathbf{F}^{t_2 \,\Arrow{.15cm}\, t_3}_{lj} \,
 -\, \sum_{k} \mathbf{F}^{t_0 \,\Arrow{.15cm}\, t_1}_{ik} \, \mathbf{F}^{t_1 \,\Arrow{.15cm}\, t_2}_{kj} \, \mathbf{F}^{t_2 \,\Arrow{.15cm}\, t_3}_{jj} \,.\label{eq:prob_union_explicit}
\end{align}
Developing the second and the third terms in Eq. (\ref{eq:prob_union_explicit}), we find:
\begin{align}\label{eq:prob_union_explicit_simple}
&P(A \cup B \cup C)_{ij} = \mathbf{F}^{t_0 \,\Arrow{.15cm}\, t_1}_{ij} 
 +\, \sum_{k\neq j} \mathbf{F}^{t_0 \,\Arrow{.15cm}\, t_1}_{ik} \, \mathbf{F}^{t_1 \,\Arrow{.15cm}\, t_2}_{kj} \,
 +\, \sum_{k \neq j} \sum_{l \neq j} \mathbf{F}^{t_0 \,\Arrow{.15cm}\, t_1}_{ik} \, \mathbf{F}^{t_1 \,\Arrow{.15cm}\, t_2}_{kl} \, \mathbf{F}^{t_2 \,\Arrow{.15cm}\, t_3}_{lj} \, = \nonumber \\
 &= \mathbf{F}^{t_0 \,\Arrow{.15cm}\, t_1}_{ij} 
 +\, \sum_{k} \mathbf{F}^{t_0 \,\Arrow{.15cm}\, t_1}_{ik} (1-\delta_{kj}) \mathbf{F}^{t_1 \,\Arrow{.15cm}\, t_2}_{kj} \,
 +\, \sum_{k, l} \mathbf{F}^{t_0 \,\Arrow{.15cm}\, t_1}_{ik} (1-\delta_{kj}) \mathbf{F}^{t_1 \,\Arrow{.15cm}\, t_2}_{kl} (1-\delta_{lj}) \mathbf{F}^{t_2 \,\Arrow{.15cm}\, t_3}_{lj} \,.\, 
\end{align}
From a geometrical point of view, impeding the indexes $k$ and $l$ from taking the value of $j$ means excluding the contribution to the union probability of all the paths that visit $j$ more than once.

Denoting with a circle the Hadamard (or element-wise) product, we can write Eq. (\ref{eq:prob_union_explicit_simple}) for any pair $i$-$j$ in a compact form and define the matrix:
\begin{equation}\label{eq:prob_union_explicit_matr}
\boldsymbol{\Gamma}^{\, f}(t_0,t_3) = \mathbf{F}^{t_0 \,\Arrow{.15cm}\, t_1} \,+\,  
\mathbf{F}^{t_0 \,\Arrow{.15cm}\, t_1} (\mathbb{L} \circ \mathbf{F}^{t_1 \,\Arrow{.15cm}\, t_2}) \,+\,  
\mathbf{F}^{t_0 \,\Arrow{.15cm}\, t_1} \Big( \mathbb{L}  \circ \big( \mathbf{F}^{t_1 \,\Arrow{.15cm}\, t_2} ( \mathbb{L} \circ  \mathbf{F}^{t_2 \,\Arrow{.15cm}\, t_3}) \big) \Big) \,,
\end{equation}
where $\mathbb{L}$ is the all-ones matrix minus the identity matrix i.e. $\mathbb{L}=\mathbb{J}-\mathbb{I}$ and $\mathbb{L}_{ij}=(1-\delta_{ij})$.

\subsubsection{Generalizing up to $M$-steps cumulated multistep explicit connectivity}
To generalize the result from the previous Section, we consider the probability of the union of $M$ different events $A_1, ..., A_M$ and, using the inclusion-exclusion formula, we can write such probability as:
\begin{align}\label{eq:inclusion-exclusion}
&P\Big( \bigcup_{i=1}^{M} A_i \Big) = \sum_{i_1=1}^{M} P(A_{i_{1}}) - \sum_{i_1 < i_2}^{M} P(A_{i_{1}} \cap A_{i_{2}}) + ...\, 
	+ (-1)^{M-1} \sum_{i_1 < ...\,<i_M}^{M} P(A_{i_{1}} \cap A_{i_{2}} \cap ...\, \cap A_{i_{M}}) = \nonumber \\ 
	&= P(A_1) + P(A_1^c \cap A_2) + P(A_1^c \cap A_2^c \cap A_3) + ... +  P(A_1^c \cap ...\cap A_{M-1}^c \cap A_{M})\,.
\end{align} 
Expanding Eq. (\ref{eq:inclusion-exclusion}) we find an expression for the probability union that is a generalization of Eq. (\ref{eq:prob_union_explicit_matr}) to the generic case of $M$-steps. Keeping fixed the initial time $t_0$, we define thus the \emph{cumulated multistep explicit connectivity} calculated forward-in-time as:
\begin{align}\label{eq:multi_int_mstepsf}
&\boldsymbol{\Gamma}^{\, f}(t_0,t_M) = \mathbf{F}^{t_0 \,\Arrow{.15cm}\, t_1} \,+\,  
\mathbf{F}^{t_0 \,\Arrow{.15cm}\, t_1} (\mathbb{L} \circ \mathbf{F}^{t_1 \,\Arrow{.15cm}\, t_2}) \,+\,  
\mathbf{F}^{t_0 \,\Arrow{.15cm}\, t_1} \Big( \mathbb{L}  \circ \big( \mathbf{F}^{t_1 \,\Arrow{.15cm}\, t_2} ( \mathbb{L} \circ  \mathbf{F}^{t_2 \,\Arrow{.15cm}\, t_3}) \big) \Big) \,+\, \nonumber \\
&\,+\, ... \,+\, \mathbf{F}^{t_0 \,\Arrow{.15cm}\, t_1} \Big( \mathbb{L}  \circ \big( \mathbf{F}^{t_1 \,\Arrow{.15cm}\, t_2}  \, ... \, ( \mathbb{L} \circ  \mathbf{F}^{t_{M-1} \,\Arrow{.15cm}\, t_M}) \, ... \, \big) \Big)\,.
\end{align}
Similarly, keeping fixed instead the final time $t_M$, we derive the \emph{cumulated multistep explicit connectivity} calculated backward-in-time:
\begin{align}\label{eq:multi_int_mstepsb}
&\boldsymbol{\Gamma}^{\, b}(t_0,t_M) =  \mathbf{B}^{t_M \,\Arrow{.15cm}\, t_{M-1}} \,+\,  
\mathbf{B}^{t_M \,\Arrow{.15cm}\, t_{M-1}} (\mathbb{L} \circ \mathbf{B}^{t_{M-1} \,\Arrow{.15cm}\, t_{M-2}}) \,+\,  
\mathbf{B}^{t_M \,\Arrow{.15cm}\, t_{M-1}} \Big( \mathbb{L}  \circ \big( \mathbf{B}^{t_{M-1} \,\Arrow{.15cm}\, t_{M-2}} ( \mathbb{L} \circ  \mathbf{B}^{t_{M-2} \,\Arrow{.15cm}\, t_{M-3}}) \big) \Big) \,+\, \nonumber \\
&\,+\, ... \,+\, \mathbf{B}^{t_M \,\Arrow{.15cm}\, t_{M-1}} \Big( \mathbb{L}  \circ \big( \mathbf{B}^{t_{M-1} \,\Arrow{.15cm}\, t_{M-2}}  \, ... \, (\mathbb{L} \circ  \mathbf{B}^{t_{1} \,\Arrow{.15cm}\, t_0}) \, ... \, \big) \Big)\,.
\end{align}

Hence, $\boldsymbol{\Gamma}^{\, f}$ and $\boldsymbol{\Gamma}^{\, b}$ provide the expected probabilities for a random walker to connect pairs of nodes in a range of possible number of steps comprised between 1 and $M$, forward- and backward-in-time respectively. Consequently, $\boldsymbol{\Gamma}^{\, f}$ corresponds also to the probability that a portion of quantity $Q$ tagged in node $i$ arrives into node $j$, forward-in-time. Similarly, $\boldsymbol{\Gamma}^{\, b}$ corresponds to the probability that a portion of sampled quantity $Q$ in node $i$ comes from node $j$, backward-in-time.

\subsubsection{Bounding $M$-steps cumulated multistep explicit connectivity probabilities}

Let's consider the forward-in-time dynamics (the same arguments hold for the backward-in-time case) and write down Eq. (\ref{eq:multi_int_mstepsf}) for a specific matrix element associated with the origin node $i$ and destination node $j$, we have:
\begin{align}
&\mathbf{F}^{t_0 \,\Arrow{.15cm}\, t_1}_{ij} 
 +\, \sum_{k_1 \neq j} \mathbf{F}^{t_0 \,\Arrow{.15cm}\, t_1}_{i k_1} \, \mathbf{F}^{t_1 \,\Arrow{.15cm}\, t_2}_{k_1 j} \,
 +\, \sum_{k_1 \neq j} \sum_{k_2 \neq j} \mathbf{F}^{t_0 \,\Arrow{.15cm}\, t_1}_{ik_1} \, \mathbf{F}^{t_1 \,\Arrow{.15cm}\, t_2}_{k_1 k_2} \, \mathbf{F}^{t_2 \,\Arrow{.15cm}\, t_3}_{k_2 j} \,
 +\ \sum_{k_1 \neq j} ... \sum_{k_{M-1} \neq j} \mathbf{F}^{t_0 \,\Arrow{.15cm}\, t_1}_{ik_1} ... \, \mathbf{F}^{t_{M-1} \,\Arrow{.15cm}\, t_M}_{k_{M-1}j} = \nonumber \\
&\mathbf{F}^{t_0 \,\Arrow{.15cm}\, t_1}_{ij} +\, \sum_{k_1 \neq j} \mathbf{F}^{t_0 \,\Arrow{.15cm}\, t_1}_{i k_1} \Big( \mathbf{F}^{t_1 \,\Arrow{.15cm}\, t_2}_{k_1 j} 
 +\, \sum_{k_2 \neq j} \mathbf{F}^{t_1 \,\Arrow{.15cm}\, t_2}_{k_1 k_2} \Big( ... \Big( \mathbf{F}^{t_{M-2} \,\Arrow{.15cm}\, t_{M-1}}_{k_{M-2}j} 
 +\, \sum_{k_{M-1} \neq j} \mathbf{F}^{t_{M-2} \,\Arrow{.15cm}\, t_{M-1}}_{k_{M-2} k_{M-1}} \mathbf{F}^{t_{M-1} \,\Arrow{.15cm}\, t_M}_{k_{M-1}j} \Big) ... \Big)\Big) \,.
\end{align}
Recalling that $\sum_j \mathbf{F}_{ij} = 1$ and $\mathbf{F}_{ij} \leq 1$, we note that the quantity in the inner parenthesis is bounded to 1. This automatically bounds to 1 the quantity in the more external parenthesis. Recursively, we can finally see that all the expression is bounded to 1 too.

\begin{figure}
    \includegraphics[width=10cm]{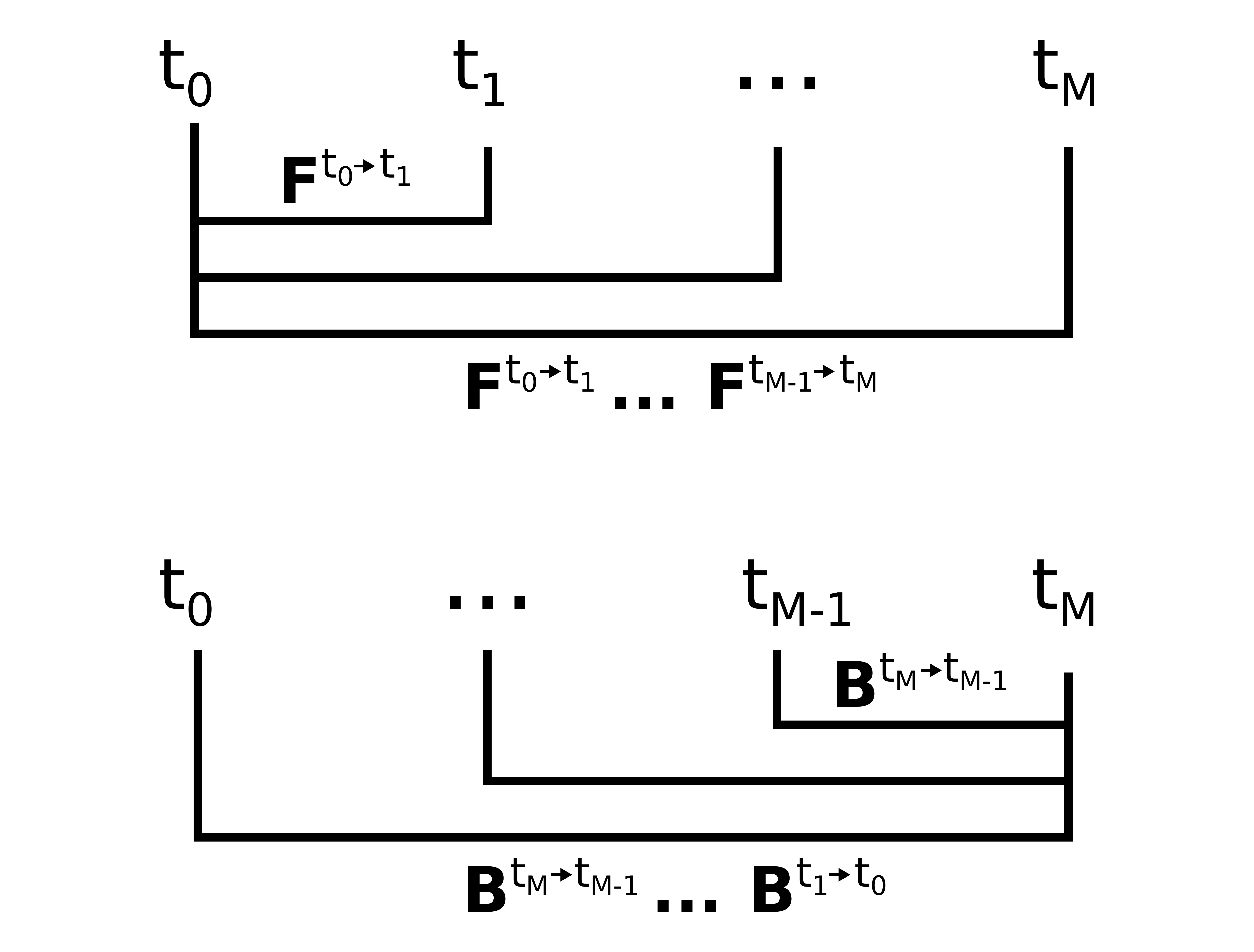} 
    \caption{Several consecutive multistep windows are used together to calculate cumulated connectivity matrices. Each window is defined by an initial time, a final time and a certain number of steps in between. One of the two times is kept fixed (either the initial or the final one) while the other is moving while it draws windows with a progressively larger number of steps. Specifically, in the forward-in-time case, the initial time $t_0$ is fixed and we increase the number of steps ending at larger $t_l$'s up to $t_M$ (top panel). Going backward-in-time, we instead end always at $t_M$ but starting from decreasing $t_l$'s until reaching $t_0$ (bottom panel).} \label{fig:multistepfig}
\end{figure}


\section{Implicit connectivity}
\label{sec:implicit}

\begin{figure}
    \includegraphics[width=10cm]{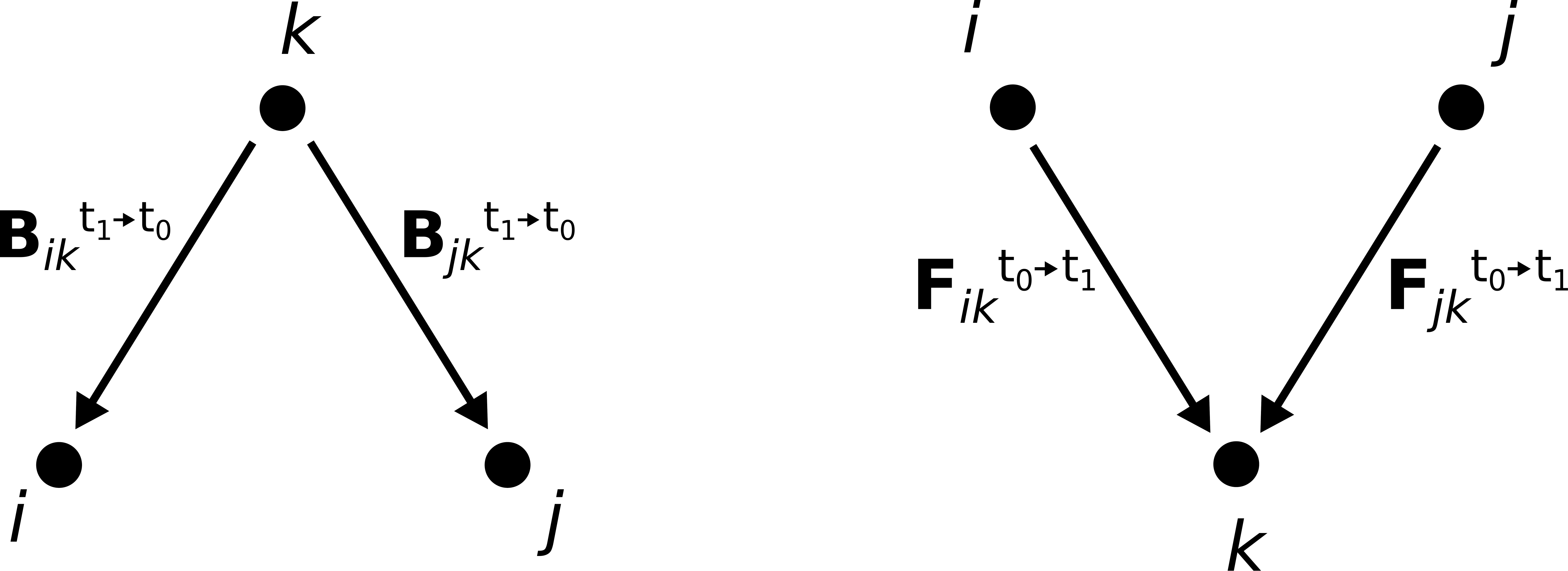}
    \caption{Schematic representation of converging (left) and diverging (right) pitchforks. Black dots represent network nodes, arrows symbolize directed temporal links.} \label{fig:vertexes}
\end{figure}

Paths are not the only connectivity patterns that can be found in a network. In general, one can identify different network motifs composed of an arbitrary number of links and nodes. Such motifs are expected to be associated with different dynamical processes depending on their geometry. In particular, we focus here on the so-called \emph{pitchforks} motifs and their associated random walk connectivity pattern that we call \emph{implicit connectivity}. We define pitchforks as a particular subgroup of motifs composed of three (sometimes two) nodes and two links. We call converging pitchfork a motif of 3 (or 2) nodes and two links pointing to one of them; we call instead diverging pitchfork a motif of 3 (or 2) nodes and two links emanated from one them (see Fig. \ref{fig:vertexes}). 

We relate such motifs to an implicit relationship between two nodes $i-j$ that are somehow influenced (or influencing) by a third node $k$. If ``third-party'' nodes $k$'s are more than one for a given pair $i-j$, we consider them together summing over $k$. The strength of these implicit relationships can be associated with the probability that two random walkers starting (or arriving) in $k$ end up (or come from) one in $i$ and the other in $j$. Similarly, the probability can be summed over $k$ to obtain the global implicit connection probability for $i-j$. As for the explicit case (Section \ref{sec:direct}), we can derive (see below) both non-cumulated and cumulated implicit connectivity probabilities and, consequently, relate these probabilities with portions of a quantity $Q$ transported across the network (see also Section \ref{subsec:transwalk}).

Note that implicit connections studied here happen ``synchronously''. For the temporal case, it means that both random walkers ensuring connections start from (or end up in) node $k$ at the same time. It is also the same time at which they reach (or start from) node $i$ and $j$, respectively. For static networks, it means that we consider for each single non-cumulated connection two random walks of the same number of steps. From a physical perspective, this is tantamount to sampling/tagging a transported quantity at the same time. This requirement is consistent with the fact that, for any dynamics running on the network, the states of each node would change in time so that it would be difficult to interpret non-synchronous relationships. More generally, if we look for a correct synoptic view of a system, we need to consider comparable snapshots of the associated network i.e. matching time intervals (temporal case) or the degrees of separation (static case). This would be the case, for instance, when studying indirect interactions between competitors for the same resource in food webs, shared ``influencers'' of opinions in social systems or common sources of pollutants in fluid transport networks.

\subsection{Pitchfork motifs and the implicit connectivity concept}

\subsubsection{Single pitchfork motifs}
From now on, let's focus on diverging pitchforks (an analogous approach can be used for the converging ones) over a time interval $[t_0;t_1]$. Both links composing the pitchfork emanated from the "source" node $k$ and point to nodes $i$ and $j$. We look for the probability that two random walkers, released simultaneously in $i$ and $j$ at $t_1$, moving backward-in-time, arrive together into $k$ at $t_0$. Such probability can be related to a sampling process on the pair $i-j$. Indeed, if we take a sample of the quantity $Q$ in $i$ at time $t_1$, the probability that such sample was in $k$ at time $t_0$ is $\mathbf{B}^{t_{1} \,\Arrow{.15cm}\, t_0}_{ik}$. Similarly, the probability for $j$ would be $\mathbf{B}^{t_{1} \,\Arrow{.15cm}\, t_0}_{jk}$. Hence, if we sample simultaneously in $i$ and $j$ at $t_1$ the probability that both samples were in $k$ at $t_0$ is:
\begin{align} \label{eq:probvert}
&\mathbf{B}^{t_{1} \,\Arrow{.15cm}\, t_0}_{ik} \mathbf{B}^{t_{1} \,\Arrow{.15cm}\, t_0}_{jk} \,.
\end{align}

\begin{figure}
    \includegraphics[width=10cm]{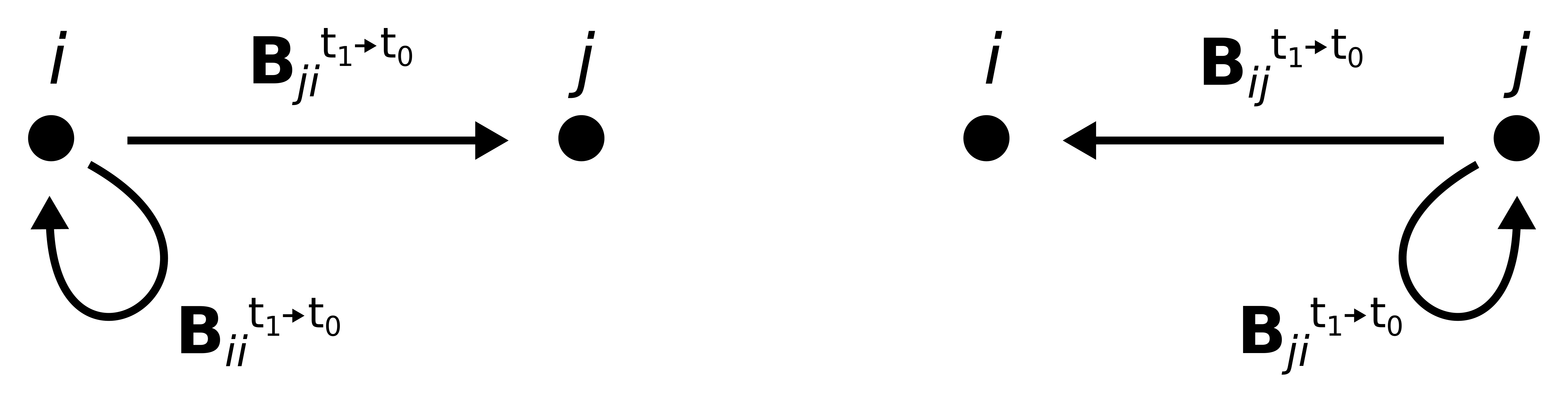}
    \caption{Degenerate pitchforks composed of two nodes instead of three i.e. when $k=i$ or $k=j$ respectively. Black dots represent network nodes, arrows symbolize directed temporal links. } \label{fig:vertexes_degen}
\end{figure}

Note that in the particular case (called here degenerate pitchfork) for which $k=i$ or $k=j$ the formulation is conceptually consistent. For instance, for $k=i$ the probability of Eq. (\ref{eq:probvert}) becomes $\mathbf{B}^{t_{1} \,\Arrow{.15cm}\, t_0}_{ii} \mathbf{B}^{t_{1} \,\Arrow{.15cm}\, t_0}_{ji}$ and the node $i$ act as source as well as destination (see Figure \ref{fig:vertexes_degen}). 

\subsubsection{Summing over pitchforks}
We now address a more general question: if one samples a quantity in nodes $i$ and $j$ at $t_1$, what is the probability that both samples share the same origin at $t_0$ (regardless of the origin nodes)? This is equivalent to looking for the probability that two random walkers, released simultaneously in $i$ and $j$ at $t_1$, arrive backward-in-time into the same node at $t_0$. By generalizing Eq. (\ref{eq:probvert}), such probability is the simple sum over all the $k$ nodes that form a pitchfork with $i$ and $j$. This is because (i) the probability that a sample in $i$ comes from $k$ is independent from the probability that a sample in $j$ comes from $k$ and because (ii) sampling quantities coming from different $k$'s inside a single node are mutually-exclusive events. We associate this backward-in-time total probability with what we call as \emph{implicit connectivity} and we define it as:
\begin{equation}\label{eq:implcon}
\mathbf{I}^{t_{1} \,\Arrow{.15cm}\, t_0}_{ij} = \mathbf{I}^{t_{1} \,\Arrow{.15cm}\, t_0}_{ji} = \sum_{k} \mathbf{B}^{t_{1} \,\Arrow{.15cm}\, t_0}_{ik} \mathbf{B}^{t_{1} \,\Arrow{.15cm}\, t_0}_{jk} = \Big( \mathbf{B}^{t_{1} \,\Arrow{.15cm}\, t_0} \transp{\mathbf{B}^{t_{1} \,\Arrow{.15cm}\, t_0}} \Big)_{ij} \,,
\end{equation} 
where with $\transp{\mathbf{B}^{t_{1} \,\Arrow{.15cm}\, t_0}}$ we denote the transpose of $\mathbf{B}^{t_{1} \,\Arrow{.15cm}\, t_0}$.

Note that when $i=j$ we have $\mathbf{I}^{t_{1} \,\Arrow{.15cm}\, t_0}_{ii} = \sum_{k} \big( \mathbf{B}^{t_{1} \,\Arrow{.15cm}\, t_0}_{ik} \big)^2$ that corresponds to the probability that two random samples of the quantity in $i$ came from the same origin (assuming a sampling with replacement). This measure corresponds to the backward-in-time Reny-entropy for $q=2$ of the node $i$ defined in \cite{ser2015flow}. Interestingly, $\mathbf{I}^{t_{1} \,\Arrow{.15cm}\, t_0}_{ii}$ is also related to the definition of the well known Simpson's Index and could be interpreted thus as a measure of diversity of origins of the quantity contained in $i$. 

For the case of converging pitchforks  an analogous development can be done. Indeed, when the two links converge to a common ``destination" node $k$, we can calculate the probability that two random walkers, released simultaneously in $i$ and $j$ at $t_0$, moving forward-in-time arrive together into $k$ at $t_1$. Such probability corresponds also to the chance that given portions of tagged quantity in $i$ and $j$ at $t_0$ will reach simultaneously $k$ at $t_1$.

\subsubsection{Bounding implicit connectivity probability}
We want to prove that $\mathbf{I}^{t_{1} \,\Arrow{.15cm}\, t_0}_{ij} \leq 1$ for every $i,j$. Using that $\mathbf{B}^{t_{1} \,\Arrow{.15cm}\, t_0}_{jk} \leq 1$ and $\mathbf{B}^{t_{1} \,\Arrow{.15cm}\, t_0}_{ik} \leq 1$ and that $\sum_{k} \mathbf{B}^{t_{1} \,\Arrow{.15cm}\, t_0}_{ik} = 1$ and $\sum_{k} \mathbf{B}^{t_{1} \,\Arrow{.15cm}\, t_0}_{jk} = 1$ one can easily find the following relationships:
\begin{align}
\mathbf{I}^{t_{1} \,\Arrow{.15cm}\, t_0}_{ij}  \leq& \sum_{k} \mathbf{B}^{t_{1} \,\Arrow{.15cm}\, t_0}_{ik} = 1 \,\\
\mathbf{I}^{t_{1} \,\Arrow{.15cm}\, t_0}_{ij}  \leq& \sum_{k} \mathbf{B}^{t_{1} \,\Arrow{.15cm}\, t_0}_{jk} = 1 \,.
\end{align}

\subsection{Non-cumulated multistep implicit connectivity}
Here, analogously to what we did for explicit connectivity (Section \ref{sec:direct}), we first define the \emph{non-cumulated multistep implicit connectivity} by focusing on a fixed number of $M$ steps (instead than single links). We develop only the case of backward implicit connectivity but the same reasoning can be used for forward-in-time dynamics. 

For $M=2$, the multistep implicit connectivity between node $i$ and $j$ is denoted as:
\begin{equation}\label{eq:implconM2}
\sum_{k} \bigg( \sum_l \mathbf{B}^{t_{2} \,\Arrow{.15cm}\, t_1}_{il} \mathbf{B}^{t_{1} \,\Arrow{.15cm}\, t_0}_{lk} 
\sum_m \mathbf{B}^{t_{2} \,\Arrow{.15cm}\, t_1}_{jm} \mathbf{B}^{t_{1} \,\Arrow{.15cm}\, t_0}_{mk} \bigg) = 
\bigg( \big( \mathbf{B}^{t_{2} \,\Arrow{.15cm}\, t_1} \mathbf{B}^{t_{1} \,\Arrow{.15cm}\, t_0} \big)
\transp{\big(} \mathbf{B}^{t_{2} \,\Arrow{.15cm}\, t_1} \mathbf{B}^{t_{1} \,\Arrow{.15cm}\, t_0} \big) \bigg)_{ij} \,.
\end{equation}
We can generalize Eq. (\ref{eq:implconM2}) formula to $M$ steps to define the \emph{non-cumulated multistep implicit connectivity} calculated forward-in-time in matrix form as:
\begin{equation}\label{eq:fimplconMgeneric}
\boldsymbol{\lambda}^{\, f}(t_0,t_M) =  \big( \mathbf{F}^{t_{0} \,\Arrow{.15cm}\, t_{1}} ...\, \mathbf{F}^{t_{M-1} \,\Arrow{.15cm}\, t_M} \big)
\transp{\big(} \mathbf{F}^{t_{0} \,\Arrow{.15cm}\, t_{1}} ...\, \mathbf{F}^{t_{M-1} \,\Arrow{.15cm}\, t_M} \big)  
\end{equation}
and the matrix form of the  \emph{non-cumulated multistep implicit connectivity} calculated backward-in-time as:
\begin{equation}\label{eq:bimplconMgeneric}
\boldsymbol{\lambda}^{\, b}(t_0,t_M) = \big( \mathbf{B}^{t_{M} \,\Arrow{.15cm}\, t_{M-1}} ...\, \mathbf{B}^{t_{1} \,\Arrow{.15cm}\, t_0} \big)
\transp{\big(} \mathbf{B}^{t_{M} \,\Arrow{.15cm}\, t_{M-1}} ...\, \mathbf{B}^{t_{1} \,\Arrow{.15cm}\, t_0} \big) .
\end{equation}

\subsection{Cumulated multistep implicit connectivity}
Similarly to Section \ref{sec:cummultiexplicit}, we now further consider the case of multiple numbers of steps to introduce the \emph{cumulated multistep implicit connectivity} (see Fig. \ref{fig:multistepfig}). We refer gain to the discrete time sequence  $\bigl\{t_0, t_{1}, \, ...\, , t_{M-1}, t_M \bigl\}$ introduced before and, without loss of generality, we consider  multistep connectivity realized in any number of steps comprised between 1 and $M$. In other words, we look for a generic analytical expression to obtain the probability of linking two nodes by implicit connections occurring over a range of possible number of steps. For Forward-in-time dynamics, the initial time $t_0$ is fixed while the number of steps considered increase successively up to largest $t_l$'s. Backward-in-time, the final time $t_M$ is fixed while the number of steps considered starts from the lowest $t_l$'s and increases successively.

\subsubsection{Deriving up to 3-steps cumulated multistep implicit connectivity}
We consider in the following the forward-in-time implicit connectivity 1and, as before,  all the derivations are similar for the backward-in-time case. Keeping fixed the initial time $t_0$, we focus on the nodes $i$ and $j$ and we want to find an expression for the union of the multistep implicit probabilities increasing progressively the number of steps from 1 to $M$. Since the probabilities at different numbers of steps are not mutually-exclusive, we can cannot use the simple probability sum. First, we evaluate the probability union from 1 to 3 steps and then we generalize it up to a generic $M$. We define the three events:
\begin{itemize}
\item $A$: taking a sample from $i$ and $j$ with the same origin in 1 step,
\item $B$: taking a sample from $i$ and $j$ with the same origin in 2 step,
\item $C$: taking a sample from $i$ and $j$ with the same origin in 3 step.
\end{itemize}
The union of the probabilities of the above three events, which we call cumulated multistep implicit connectivity, is derived from Eq. (\ref{eq:prob_union}). Following Eq. (\ref{eq:fimplconMgeneric}), we have:
\begin{align}
& P(A) = \sum_{k} \mathbf{F}^{t_{0} \,\Arrow{.15cm}\, t_1}_{ik} \mathbf{F}^{t_{0} \,\Arrow{.15cm}\, t_1}_{jk} , \label{eq:probaim}\\
& P(B) = \sum_k \bigg( \sum_l \mathbf{F}^{t_{0} \,\Arrow{.15cm}\, t_1}_{il} \mathbf{F}^{t_{1} \,\Arrow{.15cm}\, t_2}_{lk} 
\sum_m \mathbf{F}^{t_{0} \,\Arrow{.15cm}\, t_1}_{jm} \mathbf{F}^{t_{1} \,\Arrow{.15cm}\, t_2}_{mk} \bigg) , \label{eq:probbim}\\
& P(C) = \sum_k \bigg( \sum_{l,f} \mathbf{F}^{t_{0} \,\Arrow{.15cm}\, t_1}_{il} \mathbf{F}^{t_{1} \,\Arrow{.15cm}\, t_2}_{lf} \mathbf{F}^{t_{2} \,\Arrow{.15cm}\, t_3}_{fk} 
\sum_{m,g} \mathbf{F}^{t_{0} \,\Arrow{.15cm}\, t_1}_{jm} \mathbf{F}^{t_{1} \,\Arrow{.15cm}\, t_2}_{mg} \mathbf{F}^{t_{2} \,\Arrow{.15cm}\, t_3}_{gk} \bigg) .\label{eq:probcim}
\end{align}
Consequently, the remaining terms of Eq. (\ref{eq:prob_union}) are:

\begin{align}
& P(A)P(B|A) = \sum_k \bigg( \sum_l \mathbf{F}^{t_{0} \,\Arrow{.15cm}\, t_1}_{il} \mathbf{F}^{t_{1} \,\Arrow{.15cm}\, t_2}_{lk} 
\mathbf{F}^{t_{0} \,\Arrow{.15cm}\, t_1}_{jl} \mathbf{F}^{t_{1} \,\Arrow{.15cm}\, t_2}_{lk} \bigg) , \label{eq:probimplicitABA} \\
& P(B)P(C|B) = \sum_k \bigg( \sum_{l,f} \mathbf{F}^{t_{0} \,\Arrow{.15cm}\, t_1}_{il} \mathbf{F}^{t_{1} \,\Arrow{.15cm}\, t_2}_{lf} \mathbf{F}^{t_{2} \,\Arrow{.15cm}\, t_3}_{fk} 
\sum_{m} \mathbf{F}^{t_{0} \,\Arrow{.15cm}\, t_1}_{jm} \mathbf{F}^{t_{1} \,\Arrow{.15cm}\, t_2}_{mf} \mathbf{F}^{t_{2} \,\Arrow{.15cm}\, t_3}_{fk} \bigg) , \\
& P(A)P(C|A) = \sum_k \bigg( \sum_{l,f} \mathbf{F}^{t_{0} \,\Arrow{.15cm}\, t_1}_{il} \mathbf{F}^{t_{1} \,\Arrow{.15cm}\, t_2}_{lf} \mathbf{F}^{t_{2} \,\Arrow{.15cm}\, t_3}_{fk} 
\sum_{g} \mathbf{F}^{t_{0} \,\Arrow{.15cm}\, t_1}_{jl} \mathbf{F}^{t_{1} \,\Arrow{.15cm}\, t_2}_{lg} \mathbf{F}^{t_{2} \,\Arrow{.15cm}\, t_3}_{gk} \bigg) ,\\
& P(A)P(B|A)P(C|A \cap B) = \sum_k \bigg( \sum_{l,f} \mathbf{F}^{t_{0} \,\Arrow{.15cm}\, t_1}_{il} \mathbf{F}^{t_{1} \,\Arrow{.15cm}\, t_2}_{lf} \mathbf{F}^{t_{2} \,\Arrow{.15cm}\, t_3}_{fk} \mathbf{F}^{t_{0} \,\Arrow{.15cm}\, t_1}_{jl} \mathbf{F}^{t_{1} \,\Arrow{.15cm}\, t_2}_{lf} \mathbf{F}^{t_{2} \,\Arrow{.15cm}\, t_3}_{fk} \bigg) \,. \label{eq:probimplicitABACAB}
\end{align}
Developing properly the sum $\sum_m$ in Eq. (\ref{eq:probbim}) and the sums $\sum_{m,g}$ in Eq. (\ref{eq:probcim}), the contributions from Eqs. (\ref{eq:probimplicitABA}-\ref{eq:probimplicitABACAB}) cancel out inside Eq. (\ref{eq:prob_union}) and we finally find:
\begin{align}
 P(A \cup B \cup C)_{ij} & = \sum_{k} \mathbf{F}^{t_{0} \,\Arrow{.15cm}\, t_1}_{ik} \mathbf{F}^{t_{0} \,\Arrow{.15cm}\, t_1}_{jk} +
  \sum_k \bigg( \sum_l \mathbf{F}^{t_{0} \,\Arrow{.15cm}\, t_1}_{il} \mathbf{F}^{t_{1} \,\Arrow{.15cm}\, t_2}_{lk} 
  \sum_{m \neq l} \mathbf{F}^{t_{0} \,\Arrow{.15cm}\, t_1}_{jm} \mathbf{F}^{t_{1} \,\Arrow{.15cm}\, t_2}_{mk} \bigg) + \nonumber \\
& + \sum_k \bigg( \sum_{l,f} \mathbf{F}^{t_{0} \,\Arrow{.15cm}\, t_1}_{il} \mathbf{F}^{t_{1} \,\Arrow{.15cm}\, t_2}_{lf} \mathbf{F}^{t_{2} \,\Arrow{.15cm}\, t_3}_{fk} 
  \sum_{m \neq l , g \neq f} \mathbf{F}^{t_{0} \,\Arrow{.15cm}\, t_1}_{jm} \mathbf{F}^{t_{1} \,\Arrow{.15cm}\, t_2}_{mg} \mathbf{F}^{t_{2} \,\Arrow{.15cm}\, t_3}_{gk} \bigg) \,. \label{eq:union_prob_implicit3}
\end{align}
From a geometrical point of view, preventing the indexes $m$ and $g$ from taking the values of $l$ and $f$ corresponds to excluding the paths starting from $i$ and $j$ that converge to any common destination node more than once. Recalling the definition of the matrix $\mathbb{L}$ as the all-ones matrix minus the identity matrix i.e. $\mathbb{L}=\mathbb{J}-\mathbb{I}$, we can write Eq. (\ref{eq:union_prob_implicit3}) as:
\begin{align}
 P(A \cup B \cup C)_{ij} & = \sum_{k} \mathbf{F}^{t_{0} \,\Arrow{.15cm}\, t_1}_{ik} \mathbf{F}^{t_{0} \,\Arrow{.15cm}\, t_1}_{jk} +
  \sum_k \bigg( \sum_{l,m} \mathbf{F}^{t_{0} \,\Arrow{.15cm}\, t_1}_{il} \mathbf{F}^{t_{1} \,\Arrow{.15cm}\, t_2}_{lk} 
  \mathbb{L}_{lm} \mathbf{F}^{t_{0} \,\Arrow{.15cm}\, t_1}_{jm} \mathbf{F}^{t_{1} \,\Arrow{.15cm}\, t_2}_{mk} \bigg) + \nonumber \\
& + \sum_k \bigg( \sum_{l,f,m,g} \mathbf{F}^{t_{0} \,\Arrow{.15cm}\, t_1}_{il} \mathbf{F}^{t_{1} \,\Arrow{.15cm}\, t_2}_{lf} \mathbf{F}^{t_{2} \,\Arrow{.15cm}\, t_3}_{fk} 
   \mathbb{L}_{lm}  \mathbb{L}_{fg}  \mathbf{F}^{t_{0} \,\Arrow{.15cm}\, t_1}_{jm} \mathbf{F}^{t_{1} \,\Arrow{.15cm}\, t_2}_{mg} \mathbf{F}^{t_{2} \,\Arrow{.15cm}\, t_3}_{gk} \bigg) \,. \label{eq:union_prob_implicit3_bis}
\end{align}
By using the Hadamard product and performing some transpositions, we can finally find an expression for Eq. (\ref{eq:union_prob_implicit3_bis}), for every pair $i-j$, in a compact form and define the matrix:
\begin{align}
&\boldsymbol{\Lambda}^{\, b}(t_0,t_3) = \mathbf{F}^{t_{0} \,\Arrow{.15cm}\, t_1} \transp{\mathbf{F}^{t_{0} \,\Arrow{.15cm}\, t_1}} +
\mathbf{F}^{t_{0} \,\Arrow{.15cm}\, t_1} \big( \mathbb{L} \circ \big( \mathbf{F}^{t_{1} \,\Arrow{.15cm}\, t_2}  \transp{\mathbf{F}^{t_{1} \,\Arrow{.15cm}\, t_2}} \big) \big) \transp{\mathbf{F}^{t_{0} \,\Arrow{.15cm}\, t_1}} + \nonumber \\
&+ \mathbf{F}^{t_{0} \,\Arrow{.15cm}\, t_1} \Big( \mathbb{L} \circ \Big( \mathbf{F}^{t_{1} \,\Arrow{.15cm}\, t_2} \big( \mathbb{L} \circ \big(  \mathbf{F}^{t_{2} \,\Arrow{.15cm}\, t_3}  \transp{\mathbf{F}^{t_{2} \,\Arrow{.15cm}\, t_3}}   \big)\big) \transp{\mathbf{F}^{t_{1} \,\Arrow{.15cm}\, t_2}} \Big) \Big)\transp{\mathbf{F}^{t_{0} \,\Arrow{.15cm}\, t_1}} \, .
\end{align}

\subsubsection{Generalizing up to $M$-steps cumulated multistep implicit connectivity}
To generalize the result derived in the above Section, we consider the probability union of $M$ different events $A_1, ..., A_M$ using the inclusion-exclusion formula of Eq. (\ref{eq:inclusion-exclusion}). Keeping fixed the initial time $t_0$, we derive thus the \emph{cumulated multistep implicit connectivity} calculated forward-in-time:
\begin{align}
&\boldsymbol{\Lambda}^{\, f}(t_0,t_M) = \mathbf{F}^{t_{0} \,\Arrow{.15cm}\, t_1} \transp{\mathbf{F}^{t_{0} \,\Arrow{.15cm}\, t_1}} +
\mathbf{F}^{t_{0} \,\Arrow{.15cm}\, t_1} \big( \mathbb{L} \circ \big( \mathbf{F}^{t_{1} \,\Arrow{.15cm}\, t_2}  \transp{\mathbf{F}^{t_{1} \,\Arrow{.15cm}\, t_2}} \big) \big) \transp{\mathbf{F}^{t_{0} \,\Arrow{.15cm}\, t_1}} + \nonumber \\
&+ \mathbf{F}^{t_{0} \,\Arrow{.15cm}\, t_1} \Big( \mathbb{L} \circ \Big( \mathbf{F}^{t_{1} \,\Arrow{.15cm}\, t_2} \big( \mathbb{L} \circ \big(  \mathbf{F}^{t_{2} \,\Arrow{.15cm}\, t_3}  \transp{\mathbf{F}^{t_{2} \,\Arrow{.15cm}\, t_3}}   \big)\big) \transp{\mathbf{F}^{t_{1} \,\Arrow{.15cm}\, t_2}} \Big) \Big)\transp{\mathbf{F}^{t_{0} \,\Arrow{.15cm}\, t_1}} + ... \, + \nonumber \\
&+ \mathbf{F}^{t_{0} \,\Arrow{.15cm}\, t_1} \bigg( \mathbb{L} \circ \bigg( \mathbf{F}^{t_{1} \,\Arrow{.15cm}\, t_2} \Big( \mathbb{L} \circ \Big( ... \big(
 \mathbf{F}^{t_{M-1} \,\Arrow{.15cm}\, t_M}  \transp{\mathbf{F}^{t_{M-1} \,\Arrow{.15cm}\, t_M}}  
 \big) ... \Big)\Big) \transp{\mathbf{F}^{t_{1} \,\Arrow{.15cm}\, t_2}} \bigg) \bigg)\transp{\mathbf{F}^{t_{0} \,\Arrow{.15cm}\, t_1}} \,.
\end{align}
Similarly, keeping fixed instead the final time $t_M$, we derive the \emph{cumulated multistep implicit connectivity} calculated backward-in-time:
\begin{align}
&\boldsymbol{\Lambda}^{\, b}(t_0,t_M) = \mathbf{B}^{t_{M} \,\Arrow{.15cm}\, t_{M-1}} \transp{\mathbf{B}^{t_{M} \,\Arrow{.15cm}\, t_{M-1}}} +
\mathbf{B}^{t_{M} \,\Arrow{.15cm}\, t_{M-1}} \big( \mathbb{L} \circ \big( \mathbf{B}^{t_{M-1} \,\Arrow{.15cm}\, t_{M-2}}  \transp{\mathbf{B}^{t_{M-1} \,\Arrow{.15cm}\, t_{M-2}}} \big) \big) \transp{\mathbf{B}^{t_{M} \,\Arrow{.15cm}\, t_{M-1}}} + \nonumber \\
&+ \mathbf{B}^{t_{M} \,\Arrow{.15cm}\, t_{M-1}} \Big( \mathbb{L} \circ \Big( \mathbf{B}^{t_{M-1} \,\Arrow{.15cm}\, t_{M-2}} \big( \mathbb{L} \circ \big(  \mathbf{B}^{t_{M-2} \,\Arrow{.15cm}\, t_{M-3}}  \transp{\mathbf{B}^{t_{M-2} \,\Arrow{.15cm}\, t_{M-3}}}   \big)\big) \transp{\mathbf{B}^{t_{M-1} \,\Arrow{.15cm}\, t_{M-2}}} \Big) \Big)\transp{\mathbf{B}^{t_{M} \,\Arrow{.15cm}\, t_{M-1}}} + ... \, + \nonumber \\
&+ \mathbf{B}^{t_{M} \,\Arrow{.15cm}\, t_{M-1}} \bigg( \mathbb{L} \circ \bigg( \mathbf{B}^{t_{M-1} \,\Arrow{.15cm}\, t_{M-2}} \Big( \mathbb{L} \circ \Big( ... \big(
 \mathbf{B}^{t_{1} \,\Arrow{.15cm}\, t_0}  \transp{\mathbf{B}^{t_{1} \,\Arrow{.15cm}\, t_0}}  
 \big) ... \Big)\Big) \transp{\mathbf{B}^{t_{M-1} \,\Arrow{.15cm}\, t_{M-2}}} \bigg) \bigg) \transp{\mathbf{B}^{t_{M} \,\Arrow{.15cm}\, t_{M-1}}} \,.
\end{align}
Hence, $\boldsymbol{\Lambda}^{\, f}$ and $\boldsymbol{\Lambda}^{\, b}$ provide the expected probabilities for two random walkers released at the same time in two nodes of the network of arriving both into the same node in over a range of steps comprised between 1 and $M$, forward- and backward-in-time respectively. Consequently, $\boldsymbol{\Lambda}^{\, f}$  corresponds also to the probability that two portions of tagged quantity $Q$ from a node pair will arrive to the same node forward-in-time. Similarly, $\boldsymbol{\Lambda}^{\, b}$ corresponds to the probability that two portions of sampled quantity $Q$ from a node pair come from the same node backward-in-time.

\section{Example applications based on simple networks}
\label{sec:toy}

\begin{figure}
    \includegraphics[width=15cm]{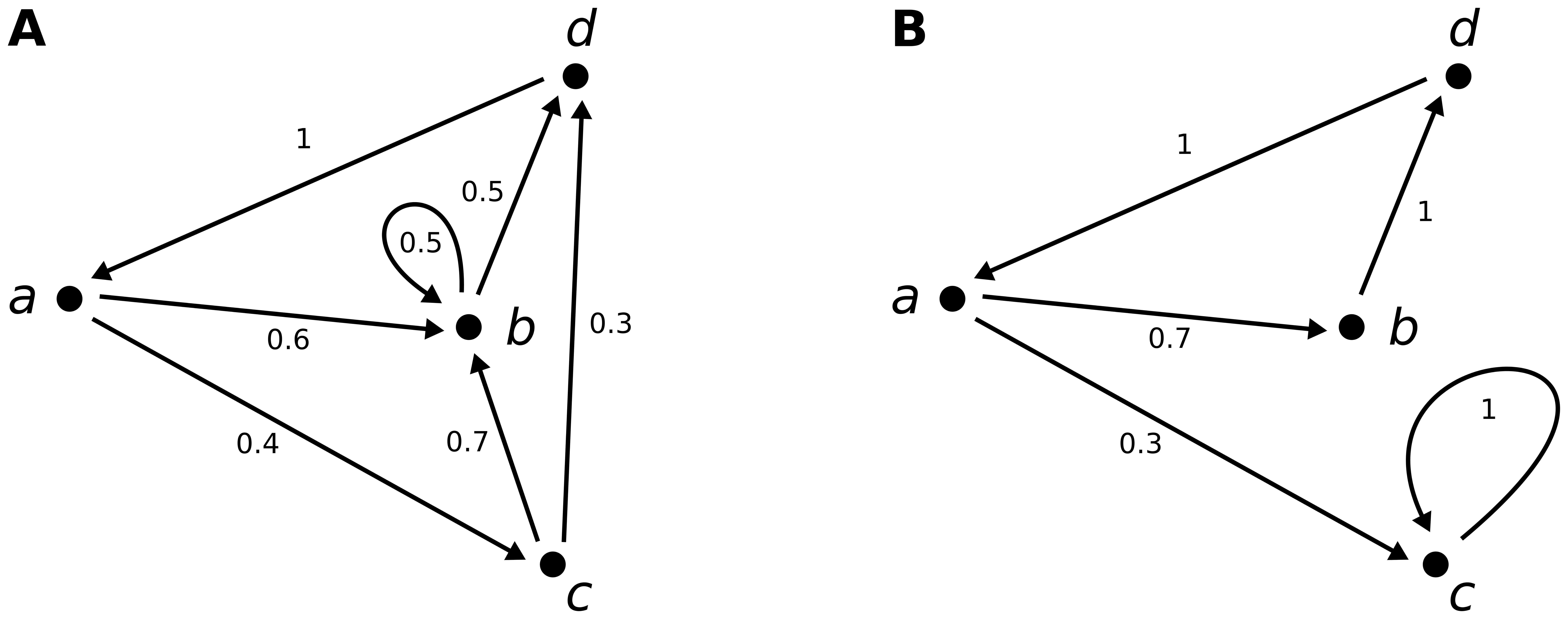}
    \caption{Examples of two small static networks:  a strongly connected network A (left panel) and a weakly connected network B (right panel). Black dots represent network nodes, arrows symbolize directed static links. Small letters label different nodes and numbers are forward-in-time probabilities of transition associated with each link. Note that while the network A is strongly connected, the network B one is not.} \label{fig:toynet}
\end{figure}

\begin{figure}
    \includegraphics[width=15cm]{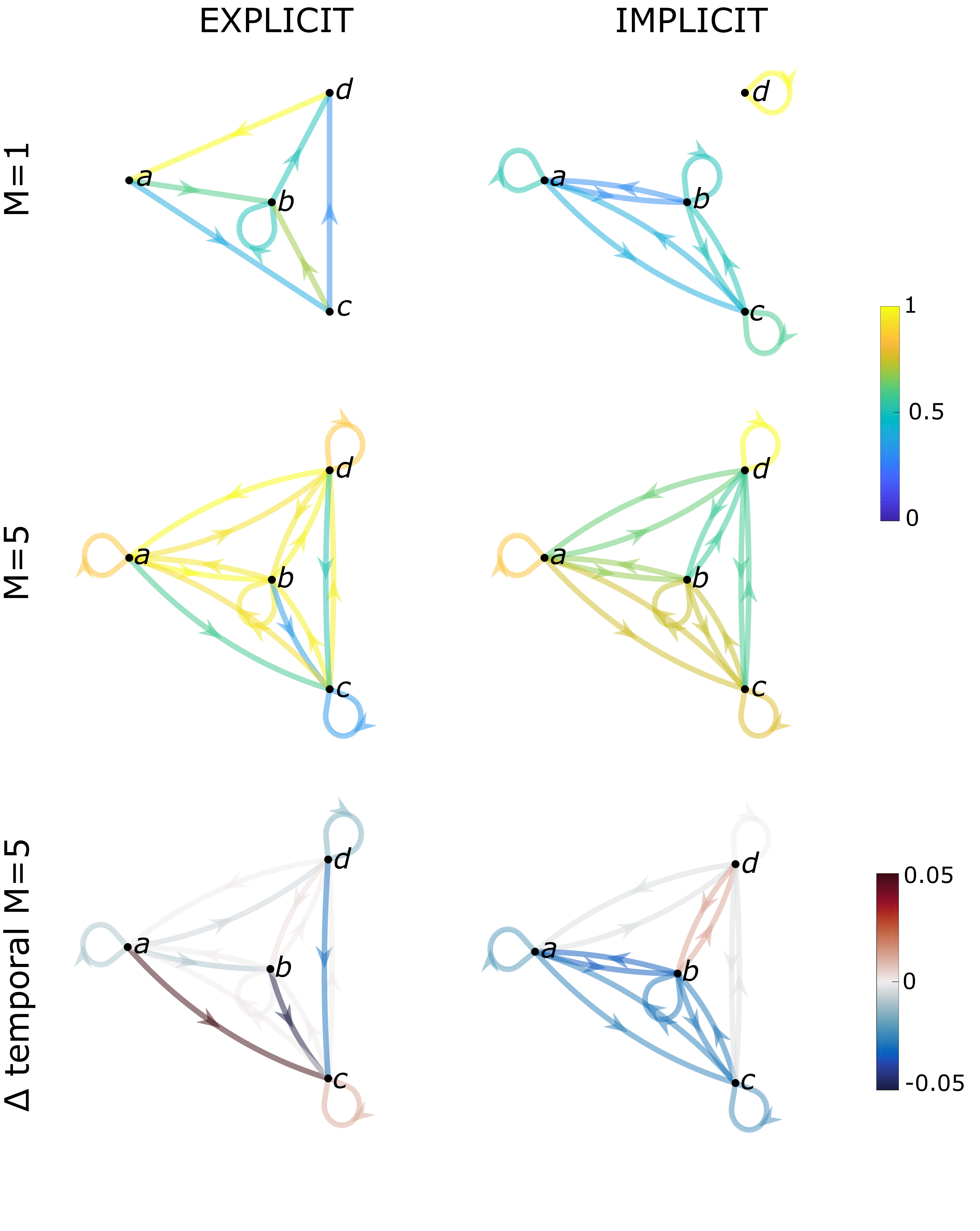}
    \caption{Schematics highlighting the differences between explicit and implicit connectivity metrics and applied to both aggregated and temporal descriptions of the network A of Fig. \ref{fig:toynet}. The four upper panels represent explicit and implicit connections for $M=1$ and $5$ with arrows colored according to their probabilities (see also Table \ref{tab:toynet}). The two lower panels shows the probability differences (i.e. the aggregated case, tab. \ref{tab:toynet}, minus the temporal case, tab. \ref{tab:toynet2}) of both explicit and implicit metrics of connectivity for $M=5$.}\label{fig:temporaldiff}
\end{figure}

We now study the performance of our novel connectivity metrics when applied on two simple, static networks represented in Fig. \ref{fig:toynet}. To start simple, we compare the analytical and the numerical results for the forward-in-time cumulated multistep explicit and implicit connectivity assuming no time-dependence (Table \ref{tab:toynet}). In Table \ref{tab:toynet} we report the values of $\boldsymbol{\Gamma^f}$ and $\boldsymbol{\Lambda^f}$ with $M=1,5$ and 100 for every pair of nodes in both networks. We then perform numerical experiments releasing thousands of random walkers across the networks verifying that their encounter probabilities match perfectly the values of $\boldsymbol{\Gamma^f}$ and $\boldsymbol{\Lambda^f}$. As highlighted in Fig \ref{fig:temporaldiff}, we clearly see that explicit and implicit connectivity present marked differences. Moreover, the results show that, while explicit connectivity is not necessarily symmetric with respect to $i$ and $j$, implicit connectivity is symmetric by definition (i.e. $\boldsymbol{\Lambda^f}_{ij} = \boldsymbol{\Lambda^f}_{ji}$).  We note also that, for explicit connectivity with $M=100$, the probabilities saturate to one only for the network A of Fig. \ref{fig:toynet}. This can be explained by the fact that the network A  is strongly connected while the network B is not, for such reason the node $c$ in the network B acts as an absorbing state for random walkers impeding the saturation to one of all probabilities.

\begin{table}
  \begin{center}
    \scalebox{0.7}{
    \begin{tabular}{l|c|c|c||c|c|c} 
            
      & \multicolumn{3}{c||}{$\boldsymbol{\Gamma^f}$}  &\multicolumn{3}{c}{$\boldsymbol{\Lambda^f$}}  \\ 
      \hline

	  & \textbf{$M=1$} & \textbf{$M=5$} & \textbf{$M=100$} & \textbf{$M=1$} & \textbf{$M=5$} & \textbf{$M=100$} \\
      \hline
	  
	  $a \rightarrow a$ & 0 & 0.855 & 1 & 0.52 & 0.84976 & 1 \\
	  \hline
	  
	  $a \rightarrow b$ & 0.6 & 0.9856 & 1 & 0.3 & 0.67607 & 1 \\
	  \hline
	  
	  $a \rightarrow c$ & 0.4 & 0.58 & 1 & 0.42 & 0.755562 & 1 \\
	  \hline
	  
	  $a \rightarrow d$ & 0 & 0.9275 & 1 & 0 & 0.6402272 & 1 \\
	  \hline
	  
	  $b \rightarrow a$ & 0 & 0.9375 & 1 & 0.3 & 0.67607 & 1 \\
	  \hline
	  
	  $b \rightarrow b$ & 0.5  & 0.94 & 1 & 0.5 & 0.73775 & 1 \\
	  \hline
	  
	  $b \rightarrow c$ & 0  & 0.35 & 1 & 0.5 & 0.73775 & 1 \\
	  \hline
	  
	  $b \rightarrow d$ & 0.5 & 0.96875 & 1 & 0 & 0.57058 & 1 \\
	  \hline
	  
	  $c \rightarrow a$ & 0 & 0.9125 & 1 & 0.42 & 0.755562 & 1 \\
	  \hline
	  
	  $c \rightarrow b$ & 0.7 & 0.964 & 1 & 0.5 & 0.73775 & 1 \\
	  \hline
	
	  $c \rightarrow c$ & 0 & 0.33 & 1 & 0.58 & 0.77971 & 1 \\
	  \hline
	  
	  $c \rightarrow d$ & 0.3 & 0.95625 & 1 & 0 & 0.58398 & 1 \\
	  \hline  
	  
	  $d \rightarrow a$ & 1 & 1.0 & 1 & 0 & 0.6402272 & 1 \\
	  \hline
	  
	  $d \rightarrow b$ & 0 & 0.952 & 1 & 0 & 0.57058 & 1 \\
	  \hline
	  
	  $d \rightarrow c$ & 0 & 0.52 & 1 & 0 & 0.58398 & 1 \\
	  \hline
	  
	  $d \rightarrow d$ & 0 & 0.855 & 1 & 1 & 1 & 1 \\
	  \hline

    \end{tabular}
    
    \quad \quad \quad \quad \quad
    
    \begin{tabular}{l|c|c|c||c|c|c} 
            
      & \multicolumn{3}{c||}{$\boldsymbol{\Gamma^f}$}  &\multicolumn{3}{c}{$\boldsymbol{\Lambda^f$}}  \\ 
      \hline

	  & \textbf{$M=1$} & \textbf{$M=5$} & \textbf{$M=100$} & \textbf{$M=1$} & \textbf{$M=5$} & \textbf{$M=100$} \\
      \hline
	  
	  $a \rightarrow a$ & 0 & 0.7 & 0.7 & 0.58 & 0.706 & 1 \\
	  \hline
	  
	  $a \rightarrow b$ & 0.7 & 0.7 & 0.7 & 0 & 0.153 & 1 \\
	  \hline
	  
	  $a \rightarrow c$ & 0.3 & 0.51 & 1 & 0.3 & 0.51 & 1 \\
	  \hline
	  
	  $a \rightarrow d$ & 0 & 0.7 & 0.7 & 0 & 0.2601 & 1 \\
	  \hline
	  
	  $b \rightarrow a$ & 0 & 1 & 1 & 0 & 0.153 & 1 \\
	  \hline
	  
	  $b \rightarrow b$ & 0 & 0.7 & 0.7 & 1 & 1 & 1 \\
	  \hline
	  
	  $b \rightarrow c$ & 0 & 0.3 & 1 & 0 & 0.3 & 1 \\
	  \hline
	  
	  $b \rightarrow d$ & 1 & 1 & 1 & 0 & 0.153 & 1 \\
	  \hline
	  
	  $c \rightarrow a$ & 0 & 0 & 0 & 0.3 & 0.51 & 1 \\
	  \hline
	  
	  $c \rightarrow b$ & 0 & 0 & 0 & 0 & 0.3 & 1 \\
	  \hline
	
	  $c \rightarrow c$ & 1 & 1 & 1 & 1 & 1 & 1 \\
	  \hline
	  
	  $c \rightarrow d$ & 0 & 0 & 0 & 0 & 0.51 & 1 \\
	  \hline
	  
	  $d \rightarrow a$ & 1 & 1.0 & 1 & 0 & 0.2601 & 1 \\
	  \hline
	  
	  $d \rightarrow b$ & 0 & 0.7 & 0.7 & 0 & 0.153 & 1 \\
	  \hline
	  
	  $d \rightarrow c$ & 0 & 0.51 & 1 & 0 & 0.51 & 1 \\
	  \hline
	  
	  $d \rightarrow d$ & 0 & 0.7 & 0.7 & 1 & 1 & 1 \\
	  \hline

    \end{tabular}
    }
   
  \end{center}
\caption{Tables reporting connectivity values of $\boldsymbol{\Gamma^f}$ and $\boldsymbol{\Lambda^f}$ with $M=1,5$ and 100 for every pair of nodes of the two example networks shown in Fig. \ref{fig:toynet}. The left table reports values for the network A and the right table reports values for the network B.}\label{tab:toynet}
\end{table}


To include the effect of temporal dynamics and highlight its relevance for connectivity patterns, we also study the case of a temporal network. To this aim, we consider the network A of Fig. \ref{fig:toynet} and we cyclically modify some of its weights while keeping the average equal to the original static network. In this way, we can properly asses the differences between a temporal network and its aggregated static counterpart. Specifically, we use the following temporal weights sequences:
\begin{align*}
a \rightarrow b \quad &: \quad (0.6, 0.5, 0.4, 0.7, 0.8) \\ 
a \rightarrow c \quad &: \quad (0.4, 0.5, 0.6, 0.3, 0.2) \\ 
c \rightarrow b \quad &: \quad (0.7, 0.6, 0.5, 0.8, 0.9) \\ 
c \rightarrow d \quad &: \quad (0.3, 0.4, 0.5, 0.2, 0.1) 
\end{align*}
where each sequence describe the weights of a link for 5 time intervals and then is repeated, the other weights are kept constant in time as in network A. In Table \ref{tab:toynet2} we report the values of $\boldsymbol{\Gamma^f}$ and $\boldsymbol{\Lambda^f}$ with $M=1,5$ and 100 for every pair of nodes in both networks. We note that, consistently, for $M=1$ and 100 the probabilities coincide with the static case. Indeed, on the one hand, in the first time interval the static network A coincide with its temporal version while, on the other hand, for $M=100$ probabilities saturate to 1 driven by links geometry rather than weight's values. As shown in Fig. \ref{fig:temporaldiff}, for $M=5$ we can find instead significant differences between the static and aggregated case that are a clear signature of the temporal dynamics.

\begin{table}
  \begin{center}
    \scalebox{0.7}{
    \begin{tabular}{l|c|c|c||c|c|c} 
            
      & \multicolumn{3}{c||}{$\boldsymbol{\Gamma^f}$}  &\multicolumn{3}{c}{$\boldsymbol{\Lambda^f$}}  \\ 
      \hline

	  & \textbf{$M=1$} & \textbf{$M=5$} & \textbf{$M=100$} & \textbf{$M=1$} & \textbf{$M=5$} & \textbf{$M=100$} \\
      \hline
	  
	  $a \rightarrow a$ & 0 & 0.865 & 1 & 0.52 & 0.8705 & 1 \\
	  \hline
	  
	  $a \rightarrow b$ & 0.6 & 0.9952 & 1 & 0.3 & 0.7102 & 1 \\
	  \hline
	  
	  $a \rightarrow c$ & 0.4 & 0.52 & 1 & 0.42 & 0.7829 & 1 \\
	  \hline
	  
	  $a \rightarrow d$ & 0 & 0.9325 & 1 & 0 & 0.6438 & 1 \\
	  \hline
	  
	  $b \rightarrow a$ & 0 & 0.9375 & 1 & 0.3 & 0.7102 & 1 \\
	  \hline
	  
	  $b \rightarrow b$ & 0.5  & 0.94 & 1 & 0.5 & 0.766 & 1 \\
	  \hline
	  
	  $b \rightarrow c$ & 0  & 0.4 & 1 & 0.5 & 0.766 & 1 \\
	  \hline
	  
	  $b \rightarrow d$ & 0.5 & 0.96875 & 1 & 0 & 0.5581 & 1 \\
	  \hline
	  
	  $c \rightarrow a$ & 0 & 0.9125 & 1 & 0.42 & 0.78296 & 1 \\
	  \hline
	  
	  $c \rightarrow b$ & 0.7 & 0.964 & 1 & 0.5 & 0.766 & 1 \\
	  \hline
	
	  $c \rightarrow c$ & 0 & 0.319 & 1 & 0.58 & 0.80344 & 1 \\
	  \hline
	  
	  $c \rightarrow d$ & 0.3 & 0.956 & 1 & 0 & 0.5869 & 1 \\
	  \hline  
	  
	  $d \rightarrow a$ & 1 & 1.0 & 1 & 0 & 0.6438 & 1 \\
	  \hline
	  
	  $d \rightarrow b$ & 0 & 0.95 & 1 & 0 & 0.5581 & 1 \\
	  \hline
	  
	  $d \rightarrow c$ & 0 & 0.55 & 1 & 0 & 0.5869 & 1 \\
	  \hline
	  
	  $d \rightarrow d$ & 0 & 0.87 & 1 & 1 & 1 & 1 \\
	  \hline

    \end{tabular}
    }
    
  \end{center}
\caption{Table reporting connectivity values of $\boldsymbol{\Gamma^f}$ and $\boldsymbol{\Lambda^f}$ with $M=1,5$ and 100 for every pair of nodes of the temporal version of the network A shown in Fig. \ref{fig:toynet}.}\label{tab:toynet2}
\end{table}

All in all, the above results suggest that $\boldsymbol{\Gamma^f}$ and $\boldsymbol{\Lambda^f}$ can provide different and complementary information about the connectivity processes occurring across a network. Moreover, as already pointed out by several study \cite{holme2012temporal}, connectivity patterns can change significantly between a full temporal network description and its aggregated counterpart and this is well reflected in our simple examples.

\section{Application to ocean transport}
\label{sec:medsea}

We now apply our theoretical framework on a real-case network representing the dynamics of fluid elements by geophysical transport processes (e.g. oceanic or atmospheric circulation). Network approaches have demonstrated great effectiveness in assessing transport and mixing of fluid parcels in both theoretical and geophysical settings \cite{ser2015flow, ser2015dominant, ser2017lagrangian, rodriguez2017clustering, lindner2017spatio, padberg2017network, wichmann2020detecting}. Studying the connectivity of such networks consists in estimating the probability of exchanging fluid parcels among different geographical locations. Since water (air, respectively) parcels carry numerous particulate and dissolved substances, connectivity is tantamount to evaluating how any almost-passive tracer is transported and dispersed by the oceanic (atmospheric, respectively) circulation. As such, relevant applications of transport networks already include studying the spread of oceanic tracers \cite{baudena2019crossroads, ser2020impact}, micro-plastics \cite{wichmann2019influence}, biological propagules \cite{rossi2014hydrodynamic,thomas2014numerical, dubois2016linking,ramesh2019small} and of atmospheric pollutants \cite{fellini2019propagation}.

Focusing on the ocean, network-based studies recently reported the presence of both preferential corridors and semi-permeable barriers of transport within realistic oceanic flows \cite{ser2015most,ser2015flow}, as documented also by alternative methods developed from Dynamical Systems Theory \cite{wiggins2005dynamical, haller2000lagrangian}. These dynamical features, which were associated with relatively persistent fronts \cite{olascoaga2006persistent} (jet-like currents \cite{budyansky2009detection}, respectively) tend to prevent (facilitate, respectively) the chaotic advection of water parcels across (within, respectively) them. Their existence determines the magnitude of connectivity among distinct oceanic sub-regions \cite{celentano2020surface} and results in the emergence of broad-scale transport patterns \cite{rossi2014hydrodynamic,miron2017lagrangian,ser2020impact}. However, this wiev of ocean connectivity has been mainly described by only considering explicit connections associated with a precise transport duration. Hence, the cumulated and implicit approaches introduced in the previous Sections can bring new insights into how different places of the ocean can be connected by water parcels dispersal. In the following, we apply our previous analytical results to provide a broader and more general perspective of the connectivity of a realistic transport network in the Mediterranean Sea. In particular, we illustrate how our new metrics allow extracting novel and relevant information (that is well explained by current oceanographic knowledge) from a state-of-the-art oceanic flow field but we by no means intend to assess the reliability of the hydrodynamical model that generated it. 

\begin{figure}
    \includegraphics[width=15cm]{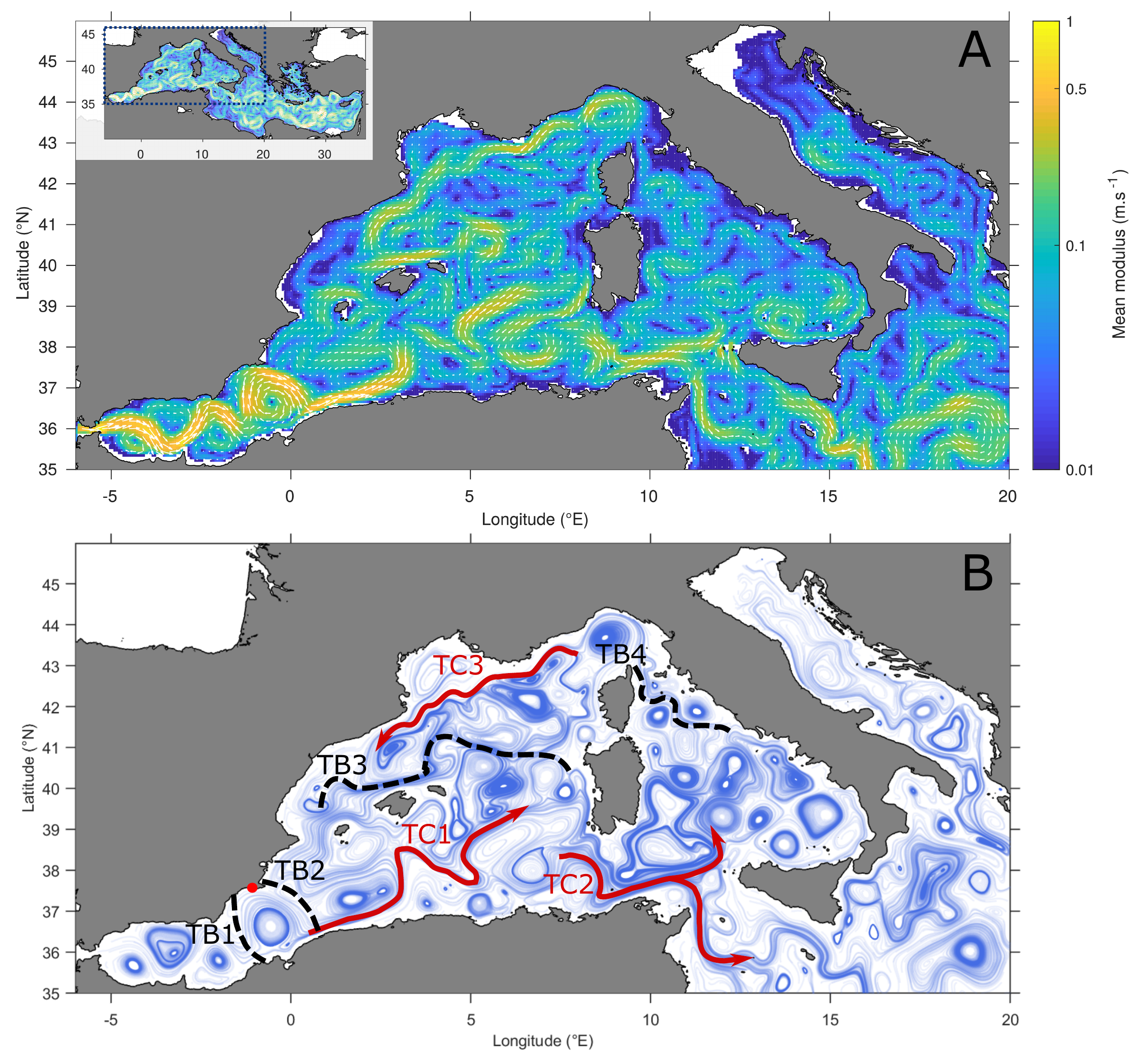}
    \caption{Maps of the study area covering the western Mediterranean Sea. Panel A: Horizontal currents direction (thin white arrows) and mean modulus (background colors, in $m s^{-1}$) of the 10 m flow field averaged over one month (01/06/2012-01/07/2012). The top-left insert displays the whole model domain covering the entire Mediterranean basin. Panel B: Streamlines of the 10 m flow field averaged over one month (01/06/2012-01/07/2012). The red dot indicates the studied coastal site located south of Cartagena. Other annotations highlight the main transport features, adapted from \cite{millot2005circulation} and \cite{poulain2012surface}. Transport Barriers (TB) are depicted in black dotted lines with the Almeria-Oran front (TB1), the Cartagena-Tenes front (TB2), the Balearic front (TB3) and the meandering barrier associated with the northern Tyrrhenian gyres (TB4). Mean positions of the preferential Transport Corridors (TC), associated with the main geostrophic jet-like currents, are represented as plain red lines with the Algerian current (TC1), the Atlantic-Ionian jet (TC2) and the Northern current (TC3).} \label{fig:myocean}
\end{figure}

Adopting the Lagrangian Flow Network approach \cite{ser2015flow}, we define a set of $N=967$ oceanic nodes representing small, equal-sized sub-regions of the Mediterranean Sea surface. Links and weights between such set of nodes quantify water parcels exchanges driven by ocean currents over a time-interval of 30 days forward-in-time. To construct the network, we use a reference horizontal flow fields produced by an operational data-assimilating ocean model (\url{//doi.org/10.25423/CMCC/MEDSEA_MULTIYEAR_PHY_006_004_E3R1}) whose outputs have been validated \cite{oddo2009nested}. More specifically, we exploit realistic daily currents at 10 m depth over a 30-day period spanning 01/06/2012 - 01/07/2012 (top-left insert in Fig. \ref{fig:myocean}A). The examination of$M$ steps on this network corresponds to the concatenation of $M$-times transport events of 30 days under the approximation of negligible diffusion and vertical displacements \cite{ser2015most,ser2015dominant}.  Explicit forward-in-time connectivity, in this case, is associated with the probability for a fluid parcel of traveling from one node to another and thus, to the probability that tagging a volume of water in one region of the ocean it will arrive to another given destination (after 30 days). Implicit connectivity represents instead the probability for two water parcels, belonging each of them to different nodes, of ending up in a third specific node. Again, this can be seen as the probability that two tagged volumes of water will meet together at a different common place in the ocean afterwards (after 30 days).

\begin{figure} 
    \includegraphics[width=18cm]{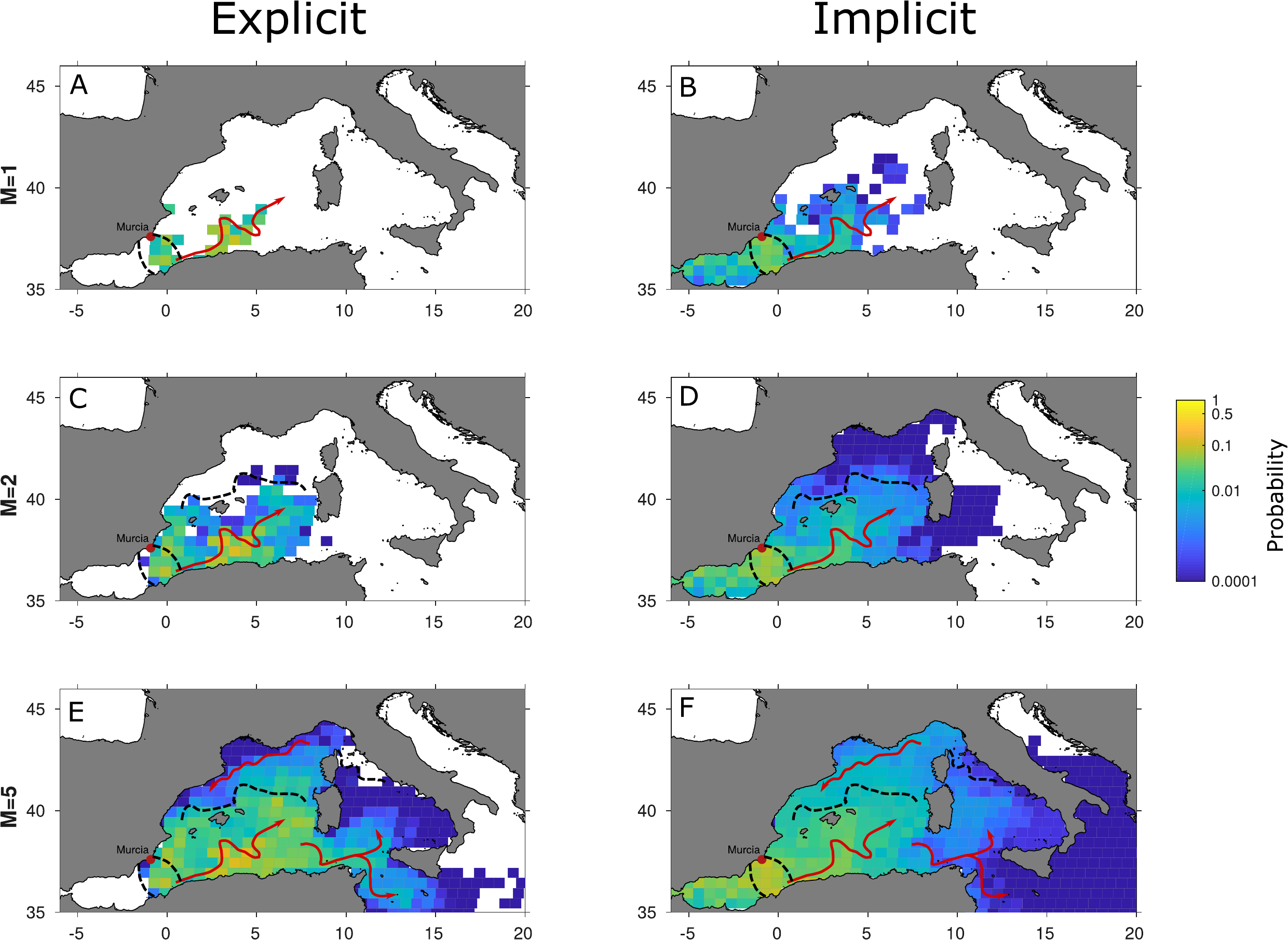}
    \caption{Forward-in-time dispersal plumes for a tracking-time of 30 days starting from Cartagena, as derived from the explicit (left panels A, C and E) and implicit (right panels B, D and F) connectivity metrics using different number of steps (from upper to lower panels: M=1, 2 and 5). Each node is colored according the probability of connection starting from our reference site (red dot) located south of Cartagena. White nodes indicate no connectivity (null probability). } \label{fig:plumemaps}
\end{figure}

We first investigate how the conventional appraisal of ocean connectivity (i.e. single step explicit) changes when computing our connectivity metrics at a few different time-steps. To do so, we arbitrarily select a coastal site located to the south of Cartagena (see the red dot in Fig. \ref{fig:myocean}B) in the Alboran Sea and we analyze the evolution of a dispersal plume starting from this reference site using both $\Gamma^{f}$ and $\Lambda^{f}$ for  $M = 1, 2$ and $5$ (Fig. \ref{fig:plumemaps}). Assigning the index $i$ to the targeted location and by considering all the non-vanishing indexes $j$, we can map all the nodes, along with their associated probabilities, which are explicitly or implicitly connected with the reference coastal site. Next, we briefly review the main transport barriers and conduits documented by previous research in the study-area and we highlight how explicit connectivity conforms with previous findings while implicit connectivity brings new insights to ocean connectivity.

In the western Mediterranean Sea, previous research highlighted, on the one hand, the presence of several Transport Barriers (TB, black annotations in Fig. \ref{fig:myocean}) associated with major oceanographic fronts: the Oran-Almeria front \cite{tintore1988study,rossi2014hydrodynamic}, the Carthagena-Tenes front \cite{hernandez2018role} and the North-balearic front \cite{mancho2006lagrangian,rossi2014hydrodynamic,ser2020impact}. On the other hand, preferential Transport Corridors (TC, red annotations in Fig. \ref{fig:myocean}) are associated with the main geostrophic jet-like currents such as the Algerian current, the Atlantic-Ionian jet and the Northern current \cite{millot2005circulation,poulain2012surface,ser2015most}.

When $M=1$ in the explicit case (Fig \ref{fig:plumemaps}) (e.g. equivalent to single-step explicit estimates), the dispersal plume is spatially inhomogeneous with two cores of medium to high probabilities ($\sim 10^{-1}$ to $10^{-2}$) which are well-explained by the pre-identified transport features (Fig \ref{fig:myocean}A, B). The Almeria-Oran and Cartagena-Tenes fronts, likely associated with an intense quasi-stationary eddy (Fig. \ref{fig:myocean}B) trap water parcels south of Cartagena while the nearshore pathway of the Algerian current (Fig. \ref{fig:myocean}B) allows some parcels to flow across TB2  and thus to disperse eastward into the Algerian basin (up to $5^{o}$E only, Fig. \ref{fig:plumemaps}). The single-step implicit connectivity plume is much larger, extending from the strait of Gibraltar to about $10^{o}$E, and associated with more homogeneous probabilities than in the explicit case. While both cores of high probabilities (ranging from $\sim 10^{-2}$ to $10^{-1}$) are similar in both cases, the implicit plume exhibits moderate to low probabilities ($\sim 10^{-3}$) in the western Alboran Sea and in the north-eastern Algerian basin. These implicit patterns are counter-intuitive and more difficult to interpret as they involve indirect connections ensured by ``third-party'' nodes. The reference site appears to be connected to the western Alboran Sea despite the presence Oran-Almeria front (TB1) and the continuous entrance of Atlantic waters (surface transport is mostly eastward) because they send waters to common downstream locations. Similarly, the low probabilities found in the north-eastern Algerian basin are probably due to recirculation processes and indirect connections ensured by coastal (counter-) currents and the meandering Algerian current. It suggests that, while the Cartagena-Tenes transport barrier constrains strongly the explicit plume \cite{hernandez2018role}, it becomes permeable in the case of the implicit plume.

For $M=2$, the explicit connectivity dispersal plume extends north-eastward, reaching $10^{o}$ E (Fig. \ref{fig:plumemaps}C). The cores of high probabilities ($\sim 10^{-1}$) match those revealed by the single-step ($M=1$) plumes, corroborating the cumulative property of our methodology. Less probable connections ($\sim 10^{-2}$ to $10^{-4}$) are found in most of the Algerian basin after approximating 60 days of advection,  whereas they were absent for $M=1$. Acting as a transport barrier, the Balearic front (TB3, Fig. \ref{fig:myocean}B) might explain why the plume does not extend further north. As such, explicit two-step connectivity suggests that our reference site remains disconnected from the French and Italian coastlines and from the Alboran Sea. Conversely, the implicit connectivity plume spreads substantially across the western Mediterranean, connecting our reference site to most shorelines until $\sim 10^{o} E$ and $\sim 45^{o} N$ (Fig. \ref{fig:plumemaps}D). Probabilities are larger or equal than $\sim 10^{-2}$ south of TB3 while they drop down to $10^{-4}$ north of the barrier (Fig. \ref{fig:plumemaps}C, D). It indicates that weak implicit connections across the Balearic front occur at $M=2$ despite the large distances and the supposed transport barrier effect. This can be explained by the fact that, in the vicinity of the front, water parcels coming from the reference site can encounter parcels coming from north of the barrier, realizing thus such implicit connections (Fig. \ref{fig:myocean}B).  

For $M=5$ (i.e. surface transport over 150 days), the multistep explicit connectivity plume (Fig. \ref{fig:plumemaps} E) spreads across most of the western Mediterranean basin and penetrates the Ionian Sea. These connectivity patterns are well-explained by the mean circulation highlighted in Fig. \ref{fig:myocean}A, B. The northern Tyrrhenian meanders (the Almeria-Oran and Cartagena-Tenes fronts, respectively) prevent effective connections with the northern Tyrrhenian Sea (the Alboran Sea, respectively). The Northern current (the Atlantic-Ionian jet, respectively) ensure rare connections ($ \sim 10^{-4}$ ; $10^{-3}$) with the northern shorelines (with the eastern Ionian Sea, respectively). The multistep implicit connectivity plume (Fig. \ref{fig:plumemaps} F) is larger: it covers the entire western Mediterranean basin and spreads over the Ionian Sea as well as the southern Adriatic Sea, despite the presence of the previously mentioned transport barriers. In comparison with the 2-step implicit connectivity, the core of elevated probabilities extends further north, suggesting that the barrier effect of the Balearic front vanishes when longer transport durations  are considered. For $M=5$ and using both explicit and implicit methods, our reference site is weakly but consistently connected to most distant coastlines, except the northern Adriatic shores.

\begin{figure}
    \includegraphics[width=15cm]{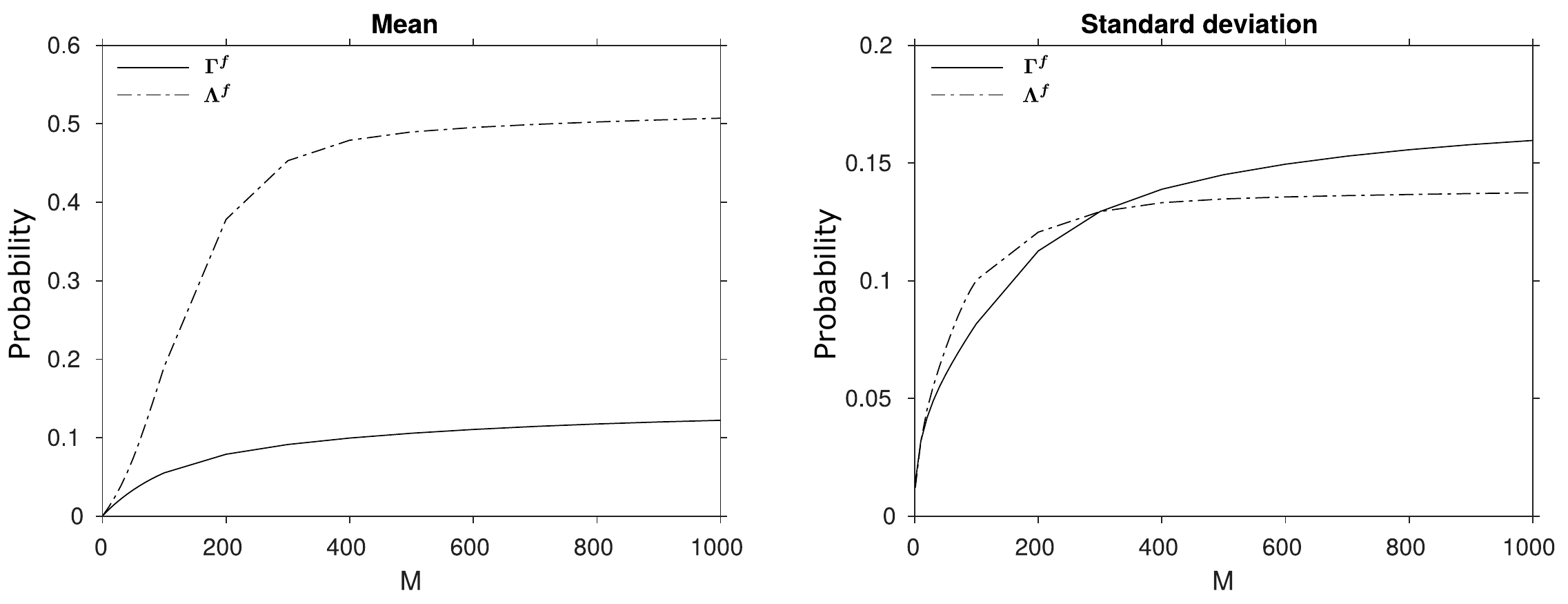}
    \caption{Mean probabilities of connection (left panel) and the associated standard deviations (right panel) for all possible pairs of nodes in our ocean network for steps ranging $M=1$ to $M=1000$. Black plain (dotted, respectively) lines stand for the cumulated multistep explicit (implicit, respectively) connectivity.} \label{fig:saturation}
\end{figure}

Finally, we analyze the global statistical distribution of our explicit and implicit proxies as a function of the number of steps. To do so, we compare the forward-in-time cumulated multistep explicit ($\Gamma^{f}$) \& implicit ($\Lambda^{f}$) connectivity metrics for different $M$ spanning $1-1000$ by computing mean probabilities of connection, and their associated standard deviations, for all pairs of nodes (i.e. $N \times N=935089$ pairs) of our flow network (Fig. \ref{fig:saturation}).We find that the mean $\Gamma^{f}$ probabilities grow sub-linearly with the number of steps until reaching a plateau at around $0.1$ after about $M=800$ steps. Mean $\Lambda^{f}$ probabilities grows almost-linearly with the number of steps until reaching a plateau at around $0.5$ after approximatively $M=500$ steps. For both metrics, saturation does not reach $1$, as it was shown for one of the theoretical cases (see Section \ref{sec:toy}), due to the presence of strongly disconnected components in our realistic ocean network. At saturation, explicit probabilities are spread-out over a wide range of values since the standard deviations tend to overcome the means. Implicit probabilities are more homogeneous and closer to the mean, even at saturation.

All in all, the newly introduced “implicit connectivity” proxy suggests thus that the connectivity of the surface ocean could have been substantially underestimated by previous methods, providing novel possible directions for the study of dispersion and transport patterns of any tracer across the ocean.

\section{Conclusions and perspectives}

Our theoretical approach can be applied to study any kind of temporal, weighted and directed network in  which a random walk can be defined. This should guarantee a broad applicability to various fields such as ecology, epidemics spreading, mobility, genetics and fluid-dynamics. By introducing the \emph{cumulated} connectivity formalism, we provide exact analytical expressions for random walk connection probabilities between any pair of nodes and across arbitrary ranges of number of steps. This framework could constitute a first step for future modeling efforts to characterize network connectivity from a probabilistic perspective. We first focused on \emph{explicit} connectivity patterns realized by paths and then for a novel \emph{implicit} connectivity concept associated with network pitchforks. Such implicit view of connectivity highlighted network topological features overlooked until now. Future studies could indeed investigate how different network topologies, such as random, small-world or scale free, would be reflected in implicit connectivity patterns and how the latter would be related to different network dynamical regimes. Moreover, when random walk single-step transition probabilities can be mapped to fractions of a given quantity exchanged across the network, it is possible to link the probabilistic connectivity interpretation to transport dynamics.  Indeed, we showed that explicit connection probabilities correspond to probabilities related to processes of tagging or sampling the transported quantity in a node forward- or backward-in-time, respectively. Analogously, implicit connection probabilities are also related to tagging or sampling processes but in two nodes simultaneously. These relationships can be further developed  both theoretically and for practical applications, such as tagging and sampling experiments on spatial systems, discovering indirect interactions in complex ecological networks or further characterize diseases spreading and opinion dynamics in social systems. Possible extensions of our approach can also include non-conservative dynamics such as production, consumption and transformation of a transported quantity by modulating the probabilities at node-scale. We finally illustrated an example of how our results can be applied to characterize fluid transport driven by ocean currents in the Mediterranean Sea. We showed that our approach extends and generalizes the way physical connectivity in the ocean was understood until now and unveils hidden connections between different regions of the Mediterranean Sea. Consequently, this changes also our understanding of the role of some oceanographic features, such as transport barriers and transport corridors, in controlling fluid connections across the seascape. Applications of this methodology to geophysical flows could provide novel insights on the spreading of drifting organisms, pollutants and, more generally, any tracer that is transported by the flow.

\bibliography{references}

\appendix

\acknowledgments

E.S-G. is grateful to C. Payrat\'{o} Borr\'{a}s for discussions on the analytical derivations. T.L. is funded by a Doctoral fellowship obtained through Aix-Marseille University. V.R. and T.L. acknowledge financial support from the European project SEAMoBB (Solutions for sEmi-Automated Monitoring of Benthic Biodiversity), funded by ERA-Net Mar-TERA (id. 145) and managed by the ANR (Grant No. ANR-17-MART-0001-02, P.I.: A. Chenuil). I. H.-C acknowledges the Vicenç Mut contract funded by the Government of the Balearic Island and the European Social Fund (ESF) Operational Programme. The Python codes used to compute connectivity probabilities are available online at \texttt{https://github.com/serjaaa/cumulated-net-conn}.

\end{document}